\documentclass[12pt,preprint,numberedappendix,twocolappendix]{emulateapj}

\newcommand\bibinc{n}		

\usepackage{subeqnarray}
\usepackage{amsmath}
\usepackage{hyperref}
\usepackage{ulem}
\usepackage{stmaryrd}

\hypersetup{colorlinks=true,linkcolor=black,citecolor=blue,urlcolor=black}

\usepackage{natbib}
\def\tt{\bf   }

\newcommand{\Eq}[1]{Equation\,(\ref{#1})}

\newcommand{\Sec}[1]{Section~\ref{#1}}

\newcommand{\Fig}[1]{Figure~\ref{#1}}

\newcommand {\twostr} {{\ttfamily TWOSTR}}
\newcommand {\disort} {{\ttfamily DISORT}}
\newcommand {\mitgcm} {{\ttfamily MITgcm}}

\def \tdrag {\tau_\mathrm{drag}}

\setcitestyle{citesep={,}}
\begin{document}

\slugcomment{Published in ApJ}

\shorttitle{Time Variability in Hot Jupiter Atmospheres}
\shortauthors{Komacek \& Showman}

\title{Temporal Variability in Hot Jupiter Atmospheres}
\author{Thaddeus D. Komacek$^{1}$, Adam P. Showman$^{2,3}$} \affil{$^1$Department of the Geophysical Sciences, The University of Chicago, Chicago, IL, 60637 \\ 
 $^2$Department of Atmospheric and Oceanic Sciences, School of Physics, Peking University, Beijing, 100871, China \\
 $^3$Lunar and Planetary Laboratory and Department of Planetary Sciences,
 University of Arizona, Tucson, AZ, 85721 \\
\url{tkomacek@uchicago.edu}} 
\begin{abstract}
Hot Jupiters receive intense incident stellar light on their daysides, which drives vigorous atmospheric circulation that attempts to erase their large dayside-to-nightside flux contrasts. Propagating waves and instabilities in hot Jupiter atmospheres can cause emergent properties of the atmosphere to be time-variable. In this work, we study such weather in hot Jupiter atmospheres using idealized cloud-free general circulation models with double-grey radiative transfer. We find that hot Jupiter atmospheres can be time-variable at the $\sim 0.1-1\%$ level in globally averaged temperature and at the $\sim 1-10\%$ level in globally averaged wind speeds. As a result, we find that observable quantities are also time variable: the secondary eclipse depth can be variable at the $\lesssim 2\%$ level, the phase curve amplitude can change by $\lesssim 1\%$, the phase curve offset can shift by $\lesssim 5^{\circ}$, and terminator-averaged wind speeds can vary by $\lesssim 2~ \mathrm{km}~\mathrm{s}^{-1}$. Additionally, we calculate how the eastern and western limb-averaged wind speeds vary with incident stellar flux and the strength of an imposed drag that parameterizes Lorentz forces in partially ionized atmospheres. We find that the eastern limb is blueshifted in models over a wide range of equilibrium temperature and drag strength, while the western limb is only redshifted if equilibrium temperatures are $\lesssim1500~\mathrm{K}$ and drag is weak. 
Lastly, we show that temporal variability may be observationally detectable in the infrared through secondary eclipse observations with \textit{JWST}, phase curve observations with future space telescopes (e.g., \textit{ARIEL}), and/or Doppler wind speed measurements with high-resolution spectrographs. 
\end{abstract}
\keywords{hydrodynamics - methods: numerical - planets and satellites: gaseous planets - planets and satellites: atmospheres}
\section{Introduction}
\label{sec:introvar}
\indent Planetary atmospheres are dynamic environments in which short-timescale temporal variability (i.e., ``weather'') is ubiquitous. Given that storms and other large-scale weather patterns are triggered by atmospheric instabilities \citep{Holton:2013}, time-variability informs us of the types of dynamical instabilities operating in planetary atmospheres. All Solar System planets experience some form of atmospheric variability \citep{Sanchez-Lavega:2010}, and so one naturally expects weather to occur in the atmospheres of exoplanets as well. In this work, we study time-variability in the atmospheres of close-in gas giant planets, or ``hot Jupiters.'' Hot Jupiters are the hottest class of exoplanets and have large atmospheric scale heights, so the observational signatures of time-variability are likely to be most apparent in this class of planet. As a result, hot Jupiters can serve as a laboratory in which to study variability using a combination of observational and theoretical approaches.  \\
\indent To date, time-variability has been observed in the atmospheres of three hot Jupiters: HAT-P-7b, Kepler-76b, and WASP-12b. \cite{Mooij2016} and \cite{Jackson:2019aa} found that the \textit{Kepler} optical light curves of HAT-P-7b and Kepler-76b show a significant variation in the phase offset, which is the temporal shift of the maximum brightness of the phase curve from secondary eclipse.  Further, the phase offset of HAT-P-7b and Kepler-76b shift from negative to positive, that is the time of maximum brightness shifts from before secondary eclipse to after secondary eclipse. Over a longer temporal baseline, \cite{Bell:2019aa} found that the \textit{Spitzer} $3.6 \ \mu\mathrm{m}$ phase curve offset of WASP-12b shifted from $32.6 \pm 6.2 ^\circ$ eastward of the substellar point in 2010 to $13.6 \pm 3.8^\circ$ westward of the substellar point in 2013. \cite{Mooij2016} and \cite{Jackson:2019aa} note that this reversal could result from a changing cloud pattern or a change in the direction of the equatorial jet in the atmospheres of these likely tidally-locked planets. Additionally, the ground-based optical observations of \cite{Essen:2019aa} found that WASP-12b's secondary eclipse depth varies on a timescale of weeks. Kepler-76b was found to have time-variability in the phase curve amplitude, while HAT-P-7b and WASP-12b did not have significant variability in amplitude. Although atmospheric circulation models of hot Jupiters that exclude magnetic effects have almost universally shown that the equatorial jet should be eastward (superrotating), it has been shown that magnetic effects due to the interaction between an intrinsic magnetic field and the thermally ionized atmosphere can reverse the equatorial jet in a time-dependent manner \citep{Rogers:2014,Hindle:2019aa}. Further, magnetic effects have been shown to qualitatively explain the phase offset variability of HAT-P-7b \citep{Rogers:2017}.  \\
\indent Additionally, significant time-variability has been observed in \textit{Spitzer} infrared secondary eclipse observations of the hot super-Earth 55 Cancri e \citep{Demory:2016,Tamburo:2018}. The mechanism inducing this variability has not been identified, but it has been hypothesized by \citeauthor{Demory:2016} to be due to volcanic plumes erupting from the molten lithosphere of the planet. \\
\indent To date, there has been no detection of time-variability in the secondary eclipse depth of hot Jupiters in the infrared. Using a sample of seven \textit{Spitzer} secondary eclipses, \cite{Agol:2010} placed an upper limit of $2.7\%$ on the variability in the secondary eclipse depth of HD 189733b at $8~\mu\mathrm{m}$. \cite{Kilpatrick:2019aa} also used \textit{Spitzer} observations to place an upper limit on the variability of HD 189733b of $5.6\%$ at $3.5\mu\mathrm{m}$ and $6.0\%$ at $4.5\mu\mathrm{m}$, and an upper limit on the variability of HD 209458b of $12\%$ at $3.5\mu\mathrm{m}$ and $1.6\%$ at $4.5\mu\mathrm{m}$. \\
\indent The observed upper limits on time-variability in the infrared secondary eclipse depth of hot Jupiters are consistent with previous modeling work. The general circulation modeling work of \cite{Showmanetal_2009} predicted small variations in the secondary eclipse depth of $\lesssim 1\%$ for HD 189733b, consistent with the observations of \cite{Agol:2010}. \cite{Heng:2011} found that the amplitude of time-variability in temperature is $\lesssim 10\%$ for HD 209458b, but did not make predictions for the variability in observable properties. Recently, \cite{Menou:2019aa} found small variability of $\lesssim 2\%$ in the dayside flux from high-resolution simulations of HD 209458b. \\
\indent There are many dynamical processes at play that induce atmospheric variability in hot Jupiter atmospheres. These processes can be broken up into two main categories: magnetohydrodynamic variability (due to the coupling of magnetism and atmospheric dynamics) and hydrodynamic variability (due to purely dynamical processes). As discussed above, magnetic effects can induce time-variability by reversing the direction of the equatorial jet. On the hottest hot Jupiters, strong electrical currents can lead to the formation of an atmospheric dynamo, which would cause significant time-variability \citep{Rogers:2017a}. Additionally, if the internally-generated dynamos of hot Jupiters are tilted with respect to the rotation axis, the flow is no longer axisymmetric, leading to significant variability in the latitudinal position of the equatorial jet \citep{Batygin:2014}. \\
\indent A wide range of purely hydrodynamic instabilities may be at play in the atmospheres of hot Jupiters. First, horizontal shear instabilities can occur due to the large difference in wind speed between the equatorial jet and its surroundings and induce barotropic eddies that sap energy from the equatorial jet \citep{Menou:2009,Fromang:2016}. Second, vertical shear instabilities 
in these stably stratified atmospheres may sap energy from the equatorial jet and limit its speed \citep{showman_2002}. Such vertical shear instabilities may cause significant energy dissipation at low pressures, resulting in variability at potentially observable levels \citep{Cho:2015,Fromang:2016}. Third, baroclinic instabilities likely play a role in driving jets in the atmospheres of gas giant planets in the Solar System \citep{Williams:1979aa,Kaspi:2006aa,Lian:2008aa,Young:2019aa}, and could cause time-variability in the atmospheres of hot Jupiters \citep{Polichtchouk:2012}. Fourth, the cloud pattern in hot Jupiter atmospheres may be patchy, and changes in this cloud pattern could lead to time-variability \citep{parmentier_2013,Mooij2016,Parmentier:2015,Lines:2018,Powell:2018aa}. Lastly, the day-night forcing could be sufficiently high amplitude such that the large-scale standing wave pattern that forces the equatorial jet becomes unstable \citep{Showman:2010}. Note that even the simulations of \cite{Liu:2013} that did not include magnetic effects or vertical shear instabilities found variations in wind speeds on the order of a few percent. \\
\indent With the high precision of the \textit{James Webb Space Telescope} (\textit{JWST}) \citep{Greene:2015}, we expect that secondary eclipse variability at the level of $\approx 1\%$ will be observable. If variability is detected, then the period and amplitude (or amplitude spectrum) of this variability can provide critical information not obtainable in any other way. In particular, the instability period and amplitude may be influenced by basic-state atmospheric properties, such as the structure of the equatorial jet and vertical stratification.
As a result, measurement of the variability properties may provide a constraint on basic-state properties of the atmosphere. \\
\indent In this paper, we place constraints on the level of observable time-variability in hot Jupiter atmospheres. To do so, we analyze the time-variability of a large suite of purely hydrodynamic simulations of the atmospheric circulation of hot Jupiters. These simulations do resolve horizontal shear instabilities, but because they are hydrostatic cannot simulate vertical shear instabilities.
Additionally, the simulations do not include magnetic effects and clouds, which could both lead to strong variability in the emitted planetary infrared flux. We also do not include the effects of hydrogen dissociation and recombination, which may cause emergent time-variability in ultra-hot Jupiter atmospheres \citep{Tan:2019aa}. As a result, one can consider our results as a baseline for understanding variability in a cloud-free and locally hydrostatic atmosphere without the effects of magnetism. This work therefore provides a crucial point of comparison for future models that study time-variability including clouds, non-hydrostatic dynamics, hydrogen dissociation and recombination, and/or magnetic effects. \\
\indent This paper is organized as follows. In \Sec{sec:methodstimevar}, we describe our numerical setup and how we diagnose time-variability in our simulations. We then go on to analyze the time-variability from our simulations, separating this time-variability into two parts. First, we analyze the physical (or ``intrinsic'') time-variability in temperature and wind speeds in \Sec{sec:resultstimevarint}. Then, in \Sec{sec:resultstimevarobs} we show the observational consequences of this intrinsic variability. We discuss the potential for observing time-variability with \textit{JWST} and ground-based high resolution spectroscopy, describe other mechanisms that may induce variability, and show model predictions for Doppler wind speeds as a function of equilibrium temperature and drag strength in \Sec{sec:disctimevar}. Lastly, we state key takeaway points in \Sec{sec:conclusionstimevar}.
\section{Numerical simulations}
\label{sec:methodstimevar}
\subsection{Numerical model}
\indent To model the time-variability in hot Jupiter atmospheres, we numerically solve the primitive equations of meteorology and represent the heating and cooling using a double-grey two-stream radiative transfer scheme. Specifically, we use the same adapted version of the three-dimensional General Circulation Model (GCM), the \mitgcm \ \citep{Adcroft:2004}, as is used in \cite{Komacek:2017,Koll:2017}, and \cite{Komacek:2019aa}. This numerical model utilizes the dynamical core of the \mitgcm~to solve the hydrostatic primitive equations of atmospheric motion on a cubed-sphere grid. Additionally, we use the \twostr~radiative transfer package \citep{Kylling:1995} of the \disort~radiative transfer code \citep{Stamnes:2027} to solve the plane-parallel radiative transfer equations with the two-stream approximation. We further use the double-grey approximation, considering only two radiative bands, one for the visible and one for the infrared. This approximation has been used previously for studies of hot Jupiter atmospheres \citep{Heng:2011a,Rauscher_2012,Komacek:2017,Tan:2019aa}, and we use the same parameterizations for the visible and infrared absorption coefficients as in \cite{Komacek:2017}. 
\subsection{Physical parameters}
\indent We use the same planetary parameters relevant for HD 209458b as in \cite{Komacek:2015} and \cite{Komacek:2017} for all simulations. We use a rotation rate of $\Omega = 2.078~\times~10^{-5}~\mathrm{s}^{-1}$, gravity of $g = 9.36~\mathrm{m}~\mathrm{s}^{-2}$, and planetary radius of $a = 9.437~\times~10^7~\mathrm{m}$. We further vary the irradiation corresponding to model equilibrium temperatures of $T_\mathrm{eq} = 500, 1000, 1500, 2000, 2500,~\mathrm{and}~3000 \ \mathrm{K}$, equivalent to varying the incident stellar flux from $1.42 \times 10^4 - 1.84 \times 10^7~\mathrm{W}~\mathrm{m}^{-2}$. We use the same absorption coefficients as in  \cite{Komacek:2017}, chosen in order to match the results of more detailed radiative transfer calculations. The absorption coefficient for incoming visible radiation is fixed at $\kappa_\mathrm{v} = 4 \times 10^{-4} \ \mathrm{m}^2\mathrm{kg}^{-1}$. The absorption coefficient for outgoing thermal radiation is set to a power law in pressure, $\kappa_\mathrm{th} = 2.28 \times 10^{-6} \left(p/1~\mathrm{Pa}\right)^{0.53} \ \mathrm{m}^2\mathrm{kg}^{-1}$. These absorption coefficients correspond to a visible photosphere that lies at a pressure of $0.23$ bars and a thermal photosphere at $0.21$ bars.  We include a small but non-negligible internal heat flux in our radiative transfer calculation. Our chosen internal heat flux is equivalent to an internal effective temperature of $T_\mathrm{int} = 100 \ \mathrm{K}$, similar to that of Jupiter. As a result, we do not include the enhancement in the internal heat flux of hot Jupiters due to internal heat deposition   \citep{Thorngren:2019aa}. \\
\indent We also impose a Rayleigh drag force that is linear in wind speed and has a magnitude inversely proportional to an assumed drag timescale $\tdrag$. As in \cite{Komacek:2017}, we consider a range of $\tdrag = 10^3, 10^4, 10^5, 10^6, 10^7,~\mathrm{and}~\infty~\mathrm{s}$. If $\tdrag \le 10^5~\mathrm{s}$, the drag timescale is spatially uniform. However, in simulations with $\tdrag > 10^5~\mathrm{s}$, we also include a basal drag term. This basal drag was first introduced by \cite{Liu:2013} in order to ensure that the equilibrated end-state of hot Jupiter simulations are independent of initial conditions and to crudely parameterize the interactions between the deep circulation and the interior of the planet. In relaxing the deep winds (near the base of the model) toward zero, we are essentially presuming that the interior has weak winds due to the weak internal heat fluxes and significant expected drag due to Lorentz forces in the convective region (e.g., \citealp{Liu:2008}). As in \cite{Komacek:2015}, we assume that this basal drag increases in strength as a power-law from no drag at a pressure of 10 bars to $\tau_\mathrm{drag,basal} = 10 \ \mathrm{days}$ at the bottom of the domain, which occurs at a pressure of 200 bars. Note that this basal drag term acts to damp atmospheric variability deep in the atmosphere (at $p\ge10~\mathrm{bars}$). We ran test simulations with $\tau_\mathrm{drag,basal} = 100 \ \mathrm{days}$ and found that the variability pattern and amplitude near the photosphere is not greatly affected by reducing the strength of basal drag. Our applied Rayleigh drag timescale is constant with time throughout all simulations.
\subsection{Numerical parameters}
\indent In our suite of numerical experiments, we use a numerical resolution of C32, which corresponds to $32 \times 32$ finite volume elements on each cube face of the cubed sphere grid and approximately equals a resolution of 128 longitudinal and 64 latitudinal grid cells. We use 40 vertical levels, 39 of which are evenly spaced in log-pressure from 200 bars to 0.2 mbars and with a top layer extending to zero pressure. These simulations include a fourth-order Shapiro filter \citep{Shapiro:1971} to damp grid-scale variations. \\
\indent In \cite{Koll:2017} it was shown that this Shapiro filter does not affect the global angular momentum budget, but can affect kinetic energy dissipation rates, especially for simulations with $\tdrag \ge 10^5 \ \mathrm{s}$. \cite{Heng:2011} found that the time-variability in GCM experiments can be affected by the strength of hyperdissipation, resolution, and the nature of the algorithm (spectral or finite-difference) used to solve the primitive equations of metorology. As a result, we computed test cases with an increased horizontal resolution of C64 and found that the amplitude of temperature and wind speed variability calculated from our nominal simulations with a horizontal resolution of C32 agree with those using a higher resolution. This agrees with the results of \cite{Menou:2019aa}, who found using a spectral dynamical core that the amplitude of variability in both wind speeds and dayside emitted flux is not strongly affected by horizontal resolution and the level of hyperdissipation. Additionally, \cite{Liu:2013} found that the circulation pattern and wind speeds are not greatly affected by increasing the horizontal resolution from C32 to C128. \cite{Koll:2017} also found that wind speeds do not significantly change when varying the horizontal resolution  and strength of the Shapiro filter. \\
\indent We run our simulations for a model time of 2,000 Earth days to spin-up the model to reach an equilibrated state in kinetic energy at observable levels (see \Sec{sec:equil} for more details). To calculate the resulting time-variability from our simulations, we output model results (e.g., temperature and wind speeds) every 12 hours for the last 150 days of the simulation. 
\subsection{Checking energy equilibration}
\label{sec:equil}
\indent As in \cite{Liu:2013}, we check the time-evolution of our simulations to determine if kinetic energy is equilibrated. We calculate the domain-integrated kinetic energy\footnote{In the primitive equations, energetic self-consistency implies that the proper expression for local kinetic energy is $\mathrm{KE} = \left(u^2 + v^2\right)/2$ and hence does not include a term for vertical velocity \citep{Holton:2013}.}, $\mathrm{KE}_\mathrm{tot}$, as  
\begin{equation}
\mathrm{KE}_\mathrm{tot} = \int_p \int_A \frac{u^2 + v^2}{2g}~dA~dp \mathrm{,}
\end{equation} 
where $u$ and $v$ are the zonal (east-west) and meridional (north-south) wind speeds at each grid point, $g = 9.36~\mathrm{m}~\mathrm{s}^{-2}$ is the assumed surface gravity, $A$ is horizontal area, and $p$ is pressure. All simulations in the suite of models reach a steady-state in domain-integrated kinetic energy by 2,000 days of model time, except weakly forced simulations with $T_\mathrm{eq} = 500 \ \mathrm{K}$ and weakly damped simulations with $\tdrag = \infty$. As a result, we do not show results from the weakly forced suite of simulations with $T_\mathrm{eq} = 500 \ \mathrm{K}$. Note that hot Jupiters are defined to have $T_\mathrm{eq} \ge 1000 \ \mathrm{K}$, so the simulations with $T_\mathrm{eq} = 500 \ \mathrm{K}$ represent tidally-locked gas giant planets that receive significantly less incident stellar radiation than hot Jupiters.
\section{Results}
\subsection{Intrinsic time-variability}
\label{sec:resultstimevarint}
\subsubsection{Temperature pattern}
\label{sec:tempvar}
\begin{figure*}
\begin{center}
\includegraphics[width=1\textwidth]{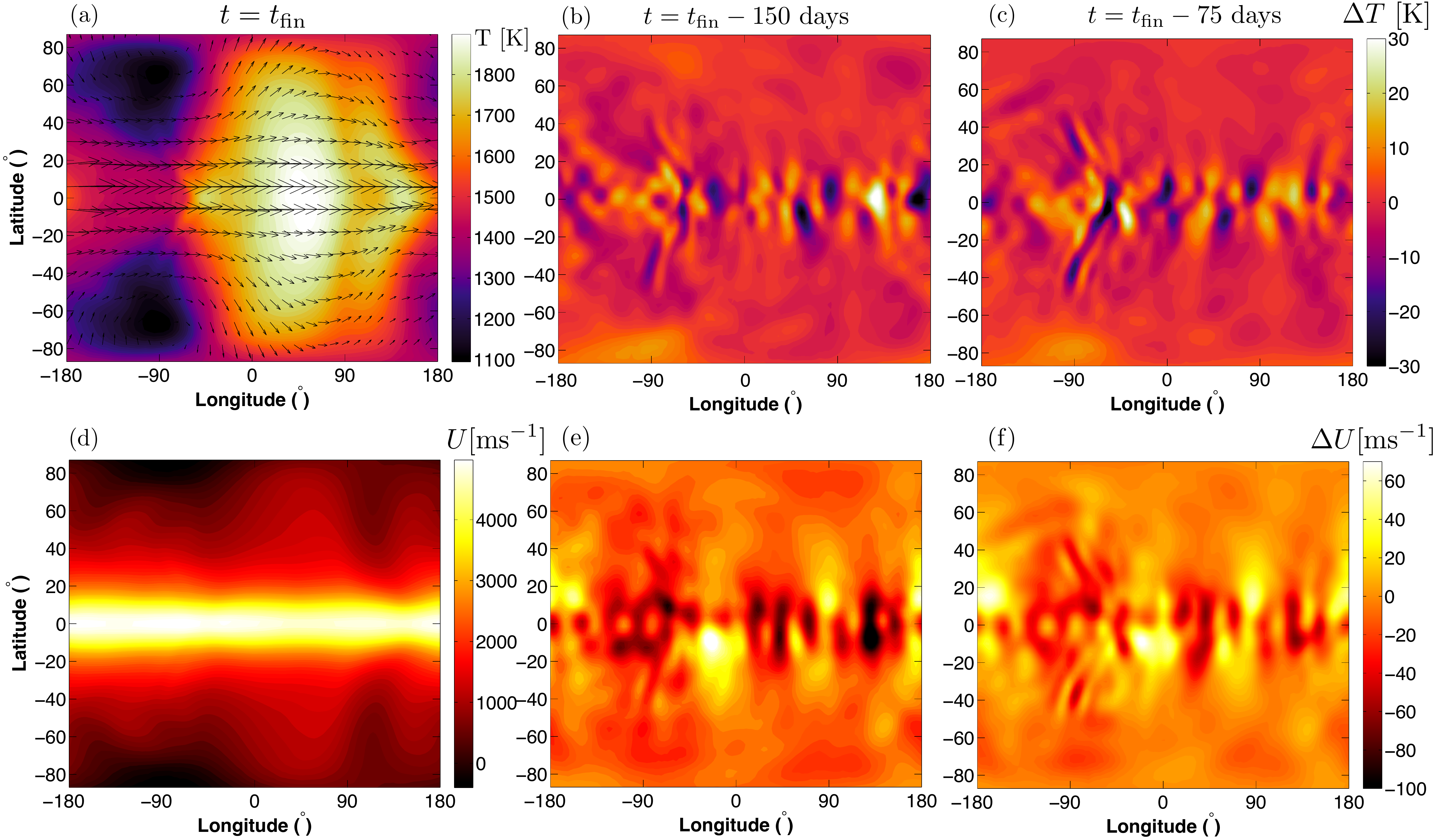}
\caption{Maps showing the typical temperature pattern and the characteristic time-variation in temperature (top) along with the typical zonal wind pattern and the characteristic time-variation in zonal wind (bottom) at 80 mbar pressure for the simulation with $T_\mathrm{eq} = 1500 \ \mathrm{K}$ and $\tau_\mathrm{drag} = \infty$ (i.e. no applied drag in the free atmosphere). (a) Temperature map with over-plotted wind vectors at the end of the simulation, $t = t_\mathrm{fin} = 2000 \ \mathrm{days}$. The characteristic superrotating jet and eastward hot spot offset of hot Jupiters is apparent. (b,c) Map of the absolute change in temperature between $t = t_\mathrm{fin} - 150 \ \mathrm{days}$ and $t = t_\mathrm{fin}$ (b) and between $t = t_\mathrm{fin} - 75 \ \mathrm{days}$ and $t = t_\mathrm{fin}$ (c). (d) Zonal wind map at the end of the simulation. (e,f) Same as (b,c) but showing maps of the absolute change in zonal wind speed. Temperature and zonal wind changes are highest in the equatorial regions and occur at the $2\%$ level or lower. 
Note that the change in temperature is largest near the western terminator, where there is strong convergence because the equatorial jet slows down as it passes the day-night terminator.}
  \label{fig:temp_maps}
 \end{center}
\end{figure*}
\begin{figure*}
\begin{center}
\includegraphics[width=1\textwidth]{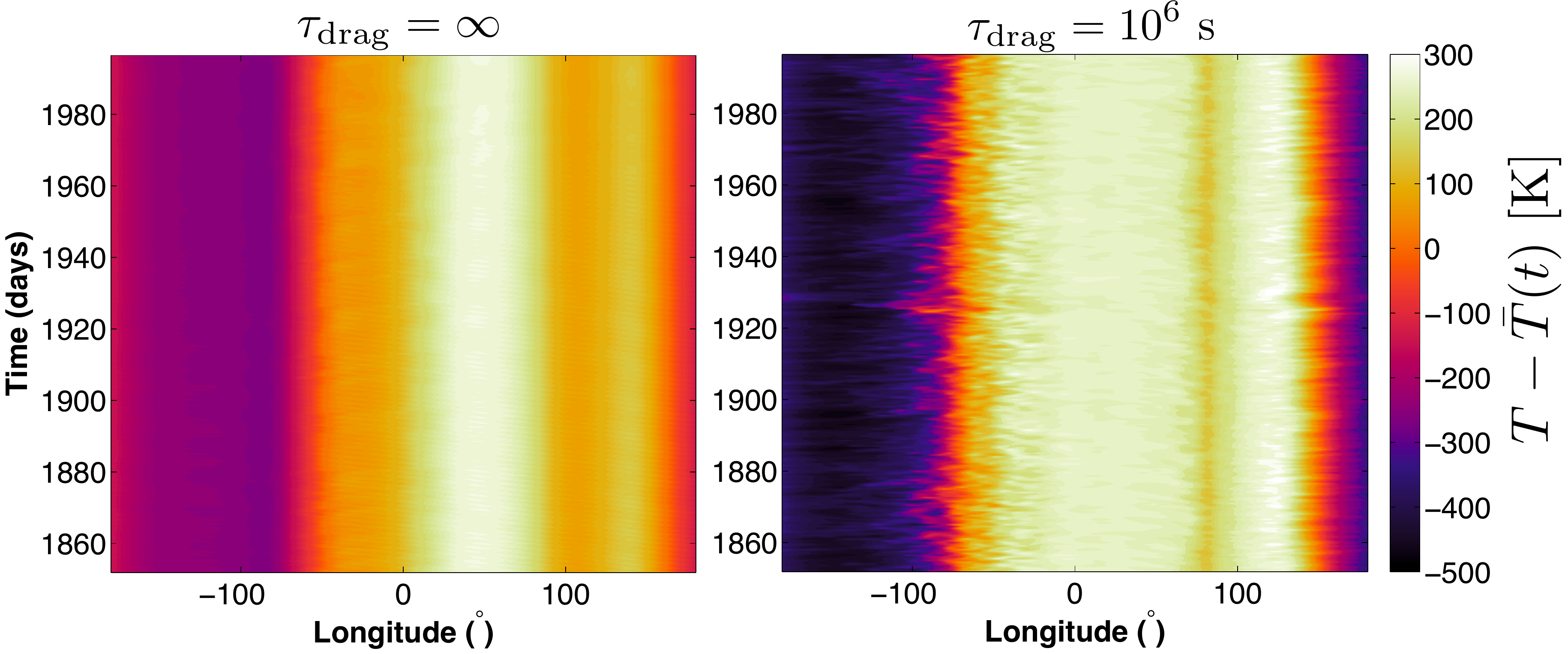}
\caption{Time-longitude section, also known as a Hovm\"{o}ller diagram, of temperature deviations from the mean longitude-averaged temperature at a given time at a pressure of $80~\mathrm{mbar}$, averaged over latitudes of $\pm 15^{\circ}$ from the equator. Shown here are results for simulations with $T_\mathrm{eq} = 1500~\mathrm{K}$ and $\tau_\mathrm{drag} = \infty~\mathrm{and}~10^6~\mathrm{s}$. Variability for both simulations is large at the western terminator (at a longitude of $-90^{\circ}$). 
The amplitude of variability is significantly larger in the case with $\tau_\mathrm{drag} = 10^6~\mathrm{s}$, partially resulting from the much larger temperature contrast at the western terminator in this simulation.}
  \label{fig:Tlongtime}
 \end{center}
\end{figure*}
\begin{figure*}
\centering
\includegraphics[width=1\textwidth]{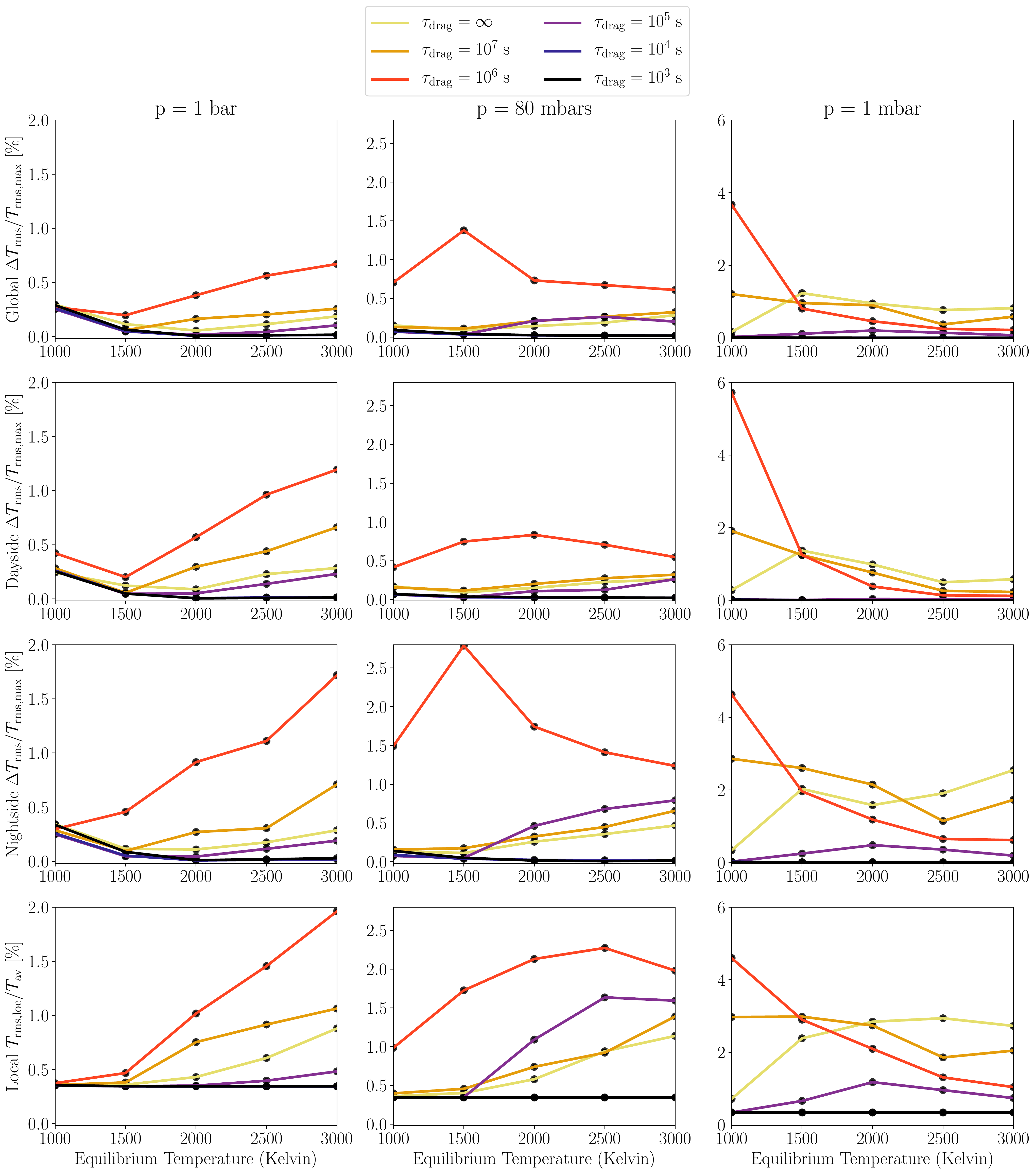}
\caption{Normalized amplitude of variability in the horizontal root-mean-square of temperature (top three rows) and locallized root-mean-square of temperature (bottom row) at given pressure levels over the last $150$ days of simulation time. The temperature variability is shown for all for simulations with varying $\tau_\mathrm{drag} = 10^3 - \infty \ \mathrm{s}$ and $T_\mathrm{eq} = 1000 - 3000 \ \mathrm{K}$ at three different pressures: $p = 1$ bar, $80$ mbars, and $1$ mbar. Plots in each column share a y-axis scale. The top row shows the global amplitude of variability, the second row shows the dayside variability amplitude, the third row shows the nightside amplitude of variability, and the bottom row shows a horizontal root-mean-square of the local measure of variability defined in \Eq{eq:trms}. The temperature variability generally increases with decreasing pressure level, and can be $\gtrsim 1\%$ at $p \leq 1 \ \mathrm{mbar}$. At pressures $\geq 80 \ \mathrm{mbar}$, the variability generally increases with increasing equilibrium temperature and is largest for the case with $\tau_\mathrm{drag} = 10^6 \ \mathrm{s}$.  At a pressure of $1 \ \mathrm{mbar}$ and $T_\mathrm{eq} \geq 1500 \ \mathrm{K}$, the global variability is largest for the case without applied drag in the free atmosphere ($\tau_\mathrm{drag} = \infty$). Cases with strong drag ($\tau_\mathrm{drag} \leq 10^5~\mathrm{s}$) generally show extremely little temperature variability. Small-scale variability (bottom row) is always present, even in simulations with strong drag.} 
  \label{fig:temp_p_var}
\end{figure*}
\indent Now we turn to analyze the intrinsic variability in hot Jupiter atmospheres, first studying the time-variability in the temperature pattern of hot Jupiters. \Fig{fig:temp_maps} (top panel) shows maps of the time-variability in temperature at a pressure of 80 mbar for a simulation with $T_\mathrm{eq} = 1500~\mathrm{K}$ and $\tdrag = \infty$, approximately representing the actual conditions of HD 209458b. Locally, there is $\lesssim 2\%$ variability over 150 days of atmospheric evolution in both temperature and zonal winds. This variability is largely confined to the equatorial regions. Notably, the equatorial regions are where the two most prominent features of hot Jupiter atmospheric circulation lie: a nearly steady equatorial planetary-scale standing wave pattern and a superrotating equatorial jet \citep{Showman_2009,Heng:2014b}. Additionally, note that the amplitude of variability is largest in localized regions near the western and eastern terminators.  \\
\indent To study in more detail how the longitudinal temperature distribution changes with time, we produce time-longitude cross sections of temperature for two of our simulations. \Fig{fig:Tlongtime} shows these Hovm\"{o}ller diagrams of temperature at $80~\mathrm{mbar}$ pressure over the last $150~\mathrm{Earth}~\mathrm{days}$ of simulation time for simulations with $T_\mathrm{eq} = 1500~\mathrm{K}$ and $\tau_\mathrm{drag} = \infty~\mathrm{and}~10^6~\mathrm{s}$. In \Fig{fig:Tlongtime}, the temperature at any given longitude is averaged between equatorial latitudes of $\pm 15^{\circ}$, where the variability is strongest. The temperature variability is significantly stronger for the case with $\tau_\mathrm{drag} = 10^6~\mathrm{s}$. \\
\indent As discussed above, the variability (which is more visible in the panel with $\tau_\mathrm{drag} = 10^6~\mathrm{s}$) is strongest at the western terminator (at a longitude of $-90^{\circ}$) and eastward of the eastern terminator (at a longitude of $\sim 130^{\circ}$). The variability at the western terminator occurs where there is a sharp temperature boundary as air coming from the cool nightside hits the hot, irradiated dayside. The feature past the eastern terminator is where air from higher levels is advected downward. This downward vertical advection results in a localized increase in the temperature relative to the cooler nightside surrounding air as a result of adiabatic heating. In general, the large-scale equatorial standing wave pattern drives this circulation \citep{Showman:2008,Rauscher:2010,Showman_Polvani_2011}, which is known as a Gill pattern in the Earth tropical dynamics literature \citep{Matsuno:1966,Gill:1980}. \\
\indent This planetary-scale equatorial standing wave (``Gill'') pattern is analogous to those in Earth's tropics in that it is comprised of equatorial Rossby and Kelvin wave modes \citep{Showman:2010,Showman_Polvani_2011,Tsai:2014,Hammond:2018aa}. These waves are triggered by the large day-to-night forcing contrast on the planet, and are damped by radiative cooling and frictional drag \citep{Perez-Becker:2013fv,Komacek:2015}. Most of the time-variability in our simulations is likely due to changes in this large-scale standing wave pattern. Though in this work we do not develop a detailed theory for the variability, we speculate on possible mechanisms in \Sec{sec:mechanism}. 
\\ \indent The time-variability in equatorial regions is significant enough that it can affect the globally averaged temperature. In \Fig{fig:temp_p_var}, we plot the variations in the root-mean-square (rms) temperature as a function of pressure, defined as
\begin{equation}
\label{eq:rmsspatial}
T_\mathrm{rms}(p) = \sqrt{\frac{\int T^2 dA}{A}} \mathrm{,}
\end{equation} 
where $A$ is horizontal area and the integral is taken on an isobar. In \Fig{fig:temp_p_var}, we show results for the integral above taken over the entire globe (top row) and over the dayside (second row) and nightside (third row) alone. We use the time-resolved output from the last 150 days of our simulation to calculate the amplitude of variability, defined as 
\begin{equation}
\label{eq:rmsmax}
\frac{\Delta T_\mathrm{rms}}{T_\mathrm{rms,max}} = \frac{T_\mathrm{rms,max} - T_\mathrm{rms,min}}{T_\mathrm{rms,max}} \mathrm{.} 
\end{equation}
In \Eq{eq:rmsmax}, $T_\mathrm{rms,max}$ and $T_\mathrm{rms,min}$ are the maximum and minimum global root-mean-square temperatures (i.e., the maximum and minimum of $T_\mathrm{rms}$) over the last 150 days of the simulation, respectively. \\
\indent It is possible that significant small-scale variability is not captured by the global-scale variability metrics discussed above. As a result, we calculate an additional metric for the root-mean-square temperature variability of local regions as a function of time over the last 150 days of the simulation:
\begin{equation}
\label{eq:trms}
T_\mathrm{rms,loc}(\lambda,\phi,p) = \sqrt{\frac{\int \left[T(\lambda,\phi,p,t) - \bar{T}(\lambda,\phi,p)\right]^2dt}{t_\mathrm{int}}} \mathrm{.}
\end{equation}
In \Eq{eq:trms}, $T_\mathrm{rms,loc}$ is the local root-mean-square temperature variability (which is a function of longitude $\lambda$, latitude $\phi$, and pressure $p$), $T$ is the spatial and time-dependent temperature, $\bar{T}$ is the time-average of the temperature over the last 150 days of simulation time, and $t_\mathrm{int} = 150~\mathrm{days}$ is the time interval over which we take the integral. Then, we calculate the spatial root-mean-square of $T_\mathrm{rms,loc}$ as in \Eq{eq:rmsspatial} to obtain a globally averaged metric for the time-variability in localized regions, and normalize by horizontal root-mean-square of the time-averaged temperature $\bar{T}$ (which we term $T_\mathrm{av}$). We show this local variability metric calculated from our suite of simulations in the bottom row of \Fig{fig:temp_p_var}. \\
\indent In general, the time-variability in global-rms temperature (top row of \Fig{fig:temp_p_var}) is of the order $0.1 - 1\%$. The local variability (bottom row of \Fig{fig:temp_p_var}) is generally larger than the global variability, and there is always significant local variability even in cases with strong drag. However, in a global-mean, much of this local variability cancels out, producing the relatively small global temperature variability. Additionally, the global variability is slightly less than the local variability in equatorial regions shown in \Fig{fig:temp_maps}, as there is much less variability at higher latitudes. The $0.1-2\%$ local temperature variability from our simulations with low $T_\mathrm{eq} \lesssim 1500~\mathrm{K}$ is similar to that found in the low-resolution finite-difference simulations of \cite{Heng:2011}, but is smaller than in their high-resolution simulations. \\
\indent The variability amplitude generally increases with decreasing pressure. This is likely because the day-to-night forcing that drives the equatorial jet increases with decreasing pressure due to the shorter radiative timescales at lower pressures \citep{Komacek:2015}. The variability amplitude on the dayside and nightside separately (second and third rows of \Fig{fig:temp_p_var}, respectively) are generally greater than in the global-average, pointing toward cancellation between the two in the global mean. For example, at any given time, hotter-than-average daysides will be accompanied by cooler-than-average nightsides. Additionally, the variability is generally larger on the nightside than on the dayside, as a similar absolute amplitude of temperature variability is relatively larger on the colder nightside. 
\subsubsection{Wind speeds}
\begin{figure*}
\begin{center}
\includegraphics[width=1\textwidth]{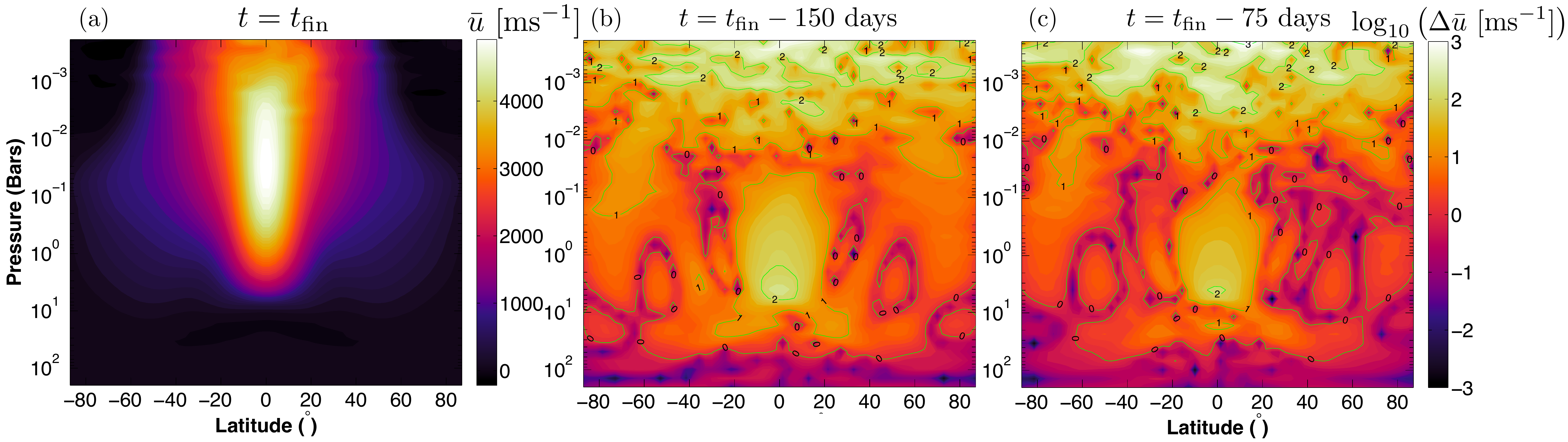}
\caption{Maps showing the typical zonal-mean zonal wind velocity and the characteristic time-variation in the zonal-mean zonal wind for the simulation with $T_\mathrm{eq} = 1500 \ \mathrm{K}$ and $\tau_\mathrm{drag} = \infty$. (a) Zonal-mean zonal wind as a function of pressure and latitude at the end of the simulation, $t = t_\mathrm{fin} = 2000 \ \mathrm{days}$. A superrotating jet, centered on the equator, is seen with a peak eastward wind speed of $\approx 4750~\mathrm{m}~\mathrm{s}^{-1}$. (b,c) Map of the logarithmic change in zonal-mean zonal wind speed between $t = t_\mathrm{fin} - 150 \ \mathrm{days}$ and $t = t_\mathrm{fin}$ (b) and between $t = t_\mathrm{fin} - 75 \ \mathrm{days}$ and $t = t_\mathrm{fin}$ (c). Contours are over-plotted in log$_{10}$ units, with contour labels of 0, 1, 2, and 3 corresponding to changes in wind speed of 1, 10, 100, and 1000 m s$^{-1}$. In the area of the superrotating jet, the change in the zonal-mean zonal wind speed is largest, with characteristic amplitudes of $\sim 1-10\%$.}
  \label{fig:wind_maps}
 \end{center}
\end{figure*}
\indent Along with the variability in temperature discussed above, our model hot Jupiter atmospheres are significantly time-variable in wind speeds. \Fig{fig:temp_maps} (bottom) shows the zonal wind speed and its variability over the last 150 days of the simulation with $T_\mathrm{eq} = 1500~\mathrm{K}$ and $\tau_\mathrm{drag} = \infty$. The zonal wind itself peaks at the equator and decreases towards higher latitudes. The variability in wind speed is also largest near the equator, with much of the variability occurring in the region where the equatorial jet is present. The location of the variability in zonal wind speed is similar to that in temperature (recall the top panel of \Fig{fig:temp_maps}), and has a similar maximum amplitude of $\sim 2\%$. \\
\indent \Fig{fig:wind_maps} shows the time-variability in the zonal-mean zonal wind speed (east-west average of the east-west wind), analogous to \Fig{fig:temp_maps} but analyzing how the wind speed variability depends on pressure. The greatest variability in zonal-mean zonal wind speed occurs near the base of the jet, at a pressure of several bars where the local variability can reach $\sim 10\%$. Note that regions closer to the expected photospheric level (shown in \Fig{fig:temp_maps}) show smaller variability, of order $1\%$ in the zonal-mean zonal wind. 
The regions on the flanks of the jet (north and south of the equatorial regions) show the least time-variability. These regions poleward of the jet are where the eddy-momentum flux that drives the equatorial jet originates \citep{Showman_Polvani_2011,Showman:2014}. \\
\begin{figure*}
\centering
\includegraphics[width=1\textwidth]{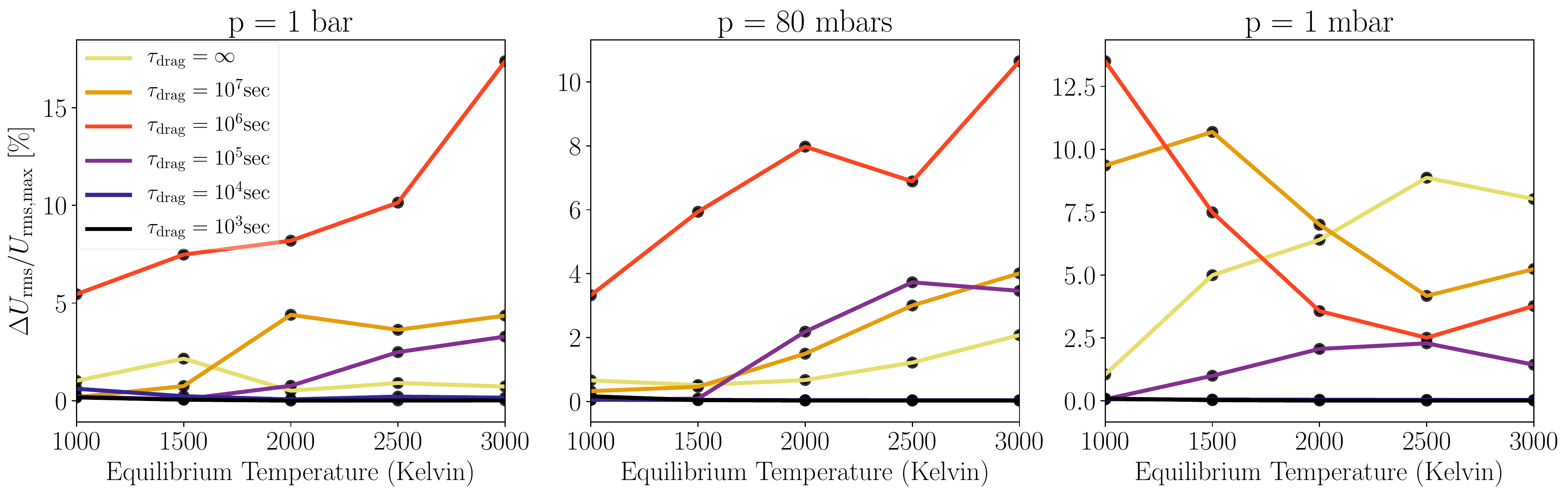}
\caption{Normalized amplitude of variability in the root-mean-square horizontal wind speed at given pressure levels over the last $150$ days of simulation time. The wind speed variability is shown for all for simulations with varying $\tau_\mathrm{drag} = 10^3 - \infty \ \mathrm{s}$ and $T_\mathrm{eq} = 1000 - 3000 \ \mathrm{K}$ at three different pressures: p = 1 bar, 80 mbars, and 1 mbar. Similar to the time-variability in the RMS of temperature (Figure \ref{fig:temp_p_var}), the variability in wind speed generally (but not always) increases with increasing equilibrium temperature and decreasing pressure level. Additionally, deep in the atmosphere ($p \geq 80~\mathrm{mbars}$), the variability is strongest for $\tau_\mathrm{drag} = 10^6 \ \mathrm{s}$.}
  \label{fig:U_p_var}
\end{figure*}
\indent The variability in wind speed discussed above manifests itself as global-scale variability with a characteristic amplitude of $1-10\%$. In \Fig{fig:U_p_var}, we show the amplitude of the rms wind speed variability at different pressures from our suite of simulations with varying equilibrium temperature and drag strength. We calculate the rms wind speed at a given pressure as 
\begin{equation}
U_\mathrm{rms}(p) = \sqrt{\frac{\int (u^2 + v^2) dA}{A}} \mathrm{,}
\end{equation}
where as before $u$ and $v$ are the zonal and meridional components of wind speed, respectively, and the integral is taken over the globe. We calculate the variability amplitude of rms wind speed as in \Eq{eq:rmsmax}, simply substituting $U_\mathrm{rms}$ for $T_\mathrm{rms}$. \\
\indent In \Fig{fig:U_p_var}, one can see that the global-rms wind speed variability is generally significantly larger than the temperature variability. As for the temperature variability, we see that the $\tdrag = 10^6~\mathrm{s}$ case exhibits relatively large variability compared to other drag timescales. 
Unlike for the temperature variability, the variability in wind speed can still be significant for $\tau_\mathrm{drag} = 10^5 \ \mathrm{s}$, and is larger than the variability for the case with $\tau_\mathrm{drag} = \infty \ \mathrm{s}$ at pressures $\ge 80~\mathrm{mbars}$ and equilibrium temperatures $\ge 2000~\mathrm{K}$. This is because the $\tau_\mathrm{drag} = 10^5 \ \mathrm{s}$ model lies between the regimes with circulation dominated by an equatorial superrotating jet and with circulation flowing from day-to-night. This leads to a forced-damped oscillation in the wind speeds, with the forcing due to the day-night irradiation difference driving wave propagation and the damping due to the applied drag. \\
\indent In general, the cases with weaker drag ($\tdrag \gtrsim 10^5~\mathrm{s}$) show much greater variability in global-rms wind speeds. In \Sec{sec:disctimevar}, we will discuss how high-resolution spectroscopic observations of hot Jupiters may be able to observe time-variability in Doppler shifts of spectral lines. 
\subsection{Observable time-variability}
\label{sec:resultstimevarobs}
\begin{figure*}
\centering
\includegraphics[width=1\textwidth]{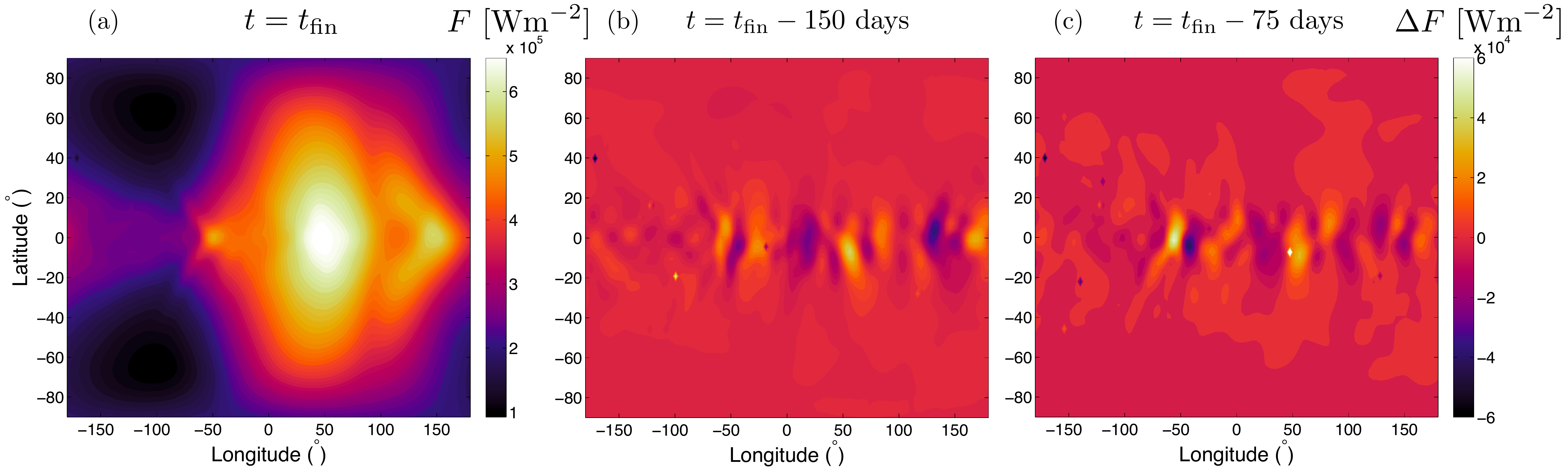}
\caption{Maps of flux at the end of the simulation and the characteristic time-variation in the emitted flux for the simulation with $T_\mathrm{eq} = 1500~\mathrm{K}$ and $\tau_\mathrm{drag} = \infty$. (a) Emitted flux at the top of the atmosphere at the end of the simulation, $t = t_\mathrm{fin} = 2000~\mathrm{days}$. The flux map closely follows the temperature map shown in \Fig{fig:temp_maps}. (b,c) Map of the change in emitted flux between $t = t_\mathrm{fin} - 150 \ \mathrm{days}$ and $t = t_\mathrm{fin}$ (b) and between $t = t_\mathrm{fin} - 75 \ \mathrm{days}$ and $t = t_\mathrm{fin}$ (c). The largest flux variations occur at low latitudes, similar to the variability in temperature and wind speed. However, the amplitude of flux variability in localized regions can be up to $\sim 10\%$, significantly larger than the local temperature variability.}
  \label{fig:fluxmapvar}
\end{figure*}
\indent To calculate how time-variability may be observable, we compute the outgoing radiated infrared flux at each time-step in each of our simulations. \Fig{fig:fluxmapvar} shows maps of the outgoing flux and its change over time for our simulation with $T_\mathrm{eq} = 1500~\mathrm{K}$ and $\tau_\mathrm{drag} = \infty$. One can see that the outgoing flux pattern looks similar to the temperature pattern shown in \Fig{fig:temp_maps}, with an eastward shift of the flux maximum and a large day-night flux contrast. Similar to the temperature variability, the flux variability is largest near the equator. However, the flux variability can be significantly larger than the temperature variability, reaching up to $\sim 10\%$ in localized regions. This is because variability in emitted flux is generally larger than that in temperature. Consider a blackbody, where $F = \sigma T^4$, with $\sigma$ the Stefan-Boltzmann constant. The relative change in flux $dF/F = 4\sigma T^3dT/F = 4\sigma T^3dT/\left(\sigma T^4\right) = 4 dT/T$. For a blackbody, the flux variability is $4$ times larger than the temperature variability. \\ 
\indent Given that we find potentially large-amplitude flux variability, we next determine to what extent this flux variability might be observable by constructing model phase curves. To do so, we calculate the flux that the Earth-facing hemisphere of the model planet radiates toward Earth at each orbital phase using the method described in Section 3.3 of \cite{Komacek:2017}. This observable flux is an area-weighted average of the flux over a hemisphere that migrates in longitude over time. 
In this way we construct a full-phase light curve, or phase curve, for each orbital period of the last 150 days in each simulation. 
\begin{figure*}
\centering
\includegraphics[width=1.0\textwidth]{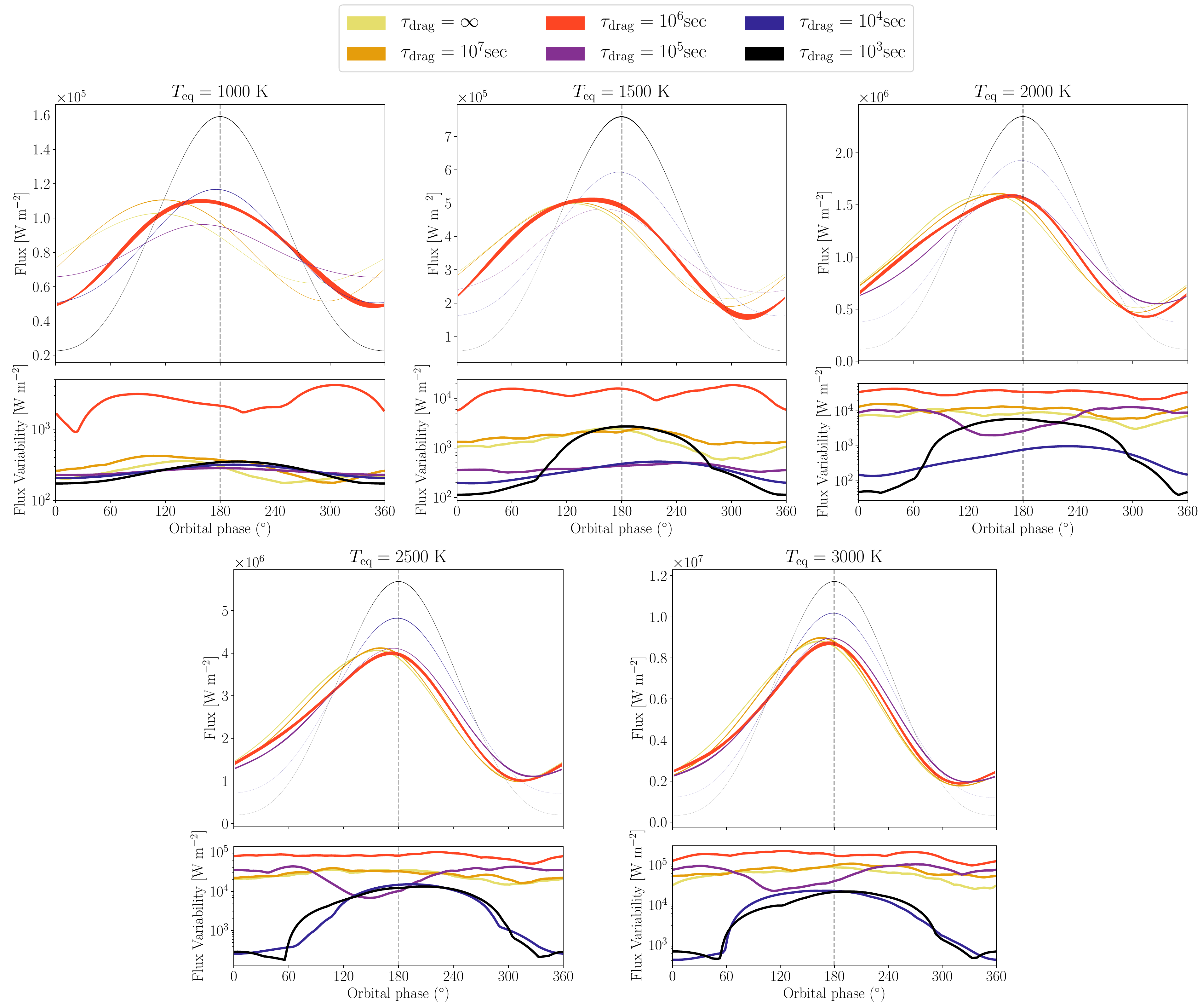}
\caption{Variability in the infrared phase curves produced by our suite of double-grey general circulation models. The top panels show the simulated emitted flux from the planet as a function of orbital phase, with the width of each line representing the maximum variation in emitted flux at each orbital phase over the last $150~\mathrm{days}$ of simulation time. The bottom panels show the maximum variability in emitted flux (the width of the lines in the top panels) as a function of orbital phase. Secondary eclipse (shown by the dashed vertical line) occurs at an orbital phase of $180^{\circ}$, and transit occurs at an orbital phase of $0^{\circ}$. Here the flux is a hemispheric average of the flux emitted toward the line of sight of the observer.  As seen for the variability in intrinsic physical parameters (see Figures \ref{fig:temp_p_var} and \ref{fig:U_p_var}), the light curve variability is largest for the case with $\tau_\mathrm{drag} = 10^6~\mathrm{s}$ and smallest in the cases with strong drag, $\tau_\mathrm{drag} \lesssim 10^4~\mathrm{s}$. Additionally, the variability is largest near secondary eclipse, due to significant variability in the phase curve offset (see \Fig{fig:lightcurve_var}).}
  \label{fig:lightcurvemaxmin}
\end{figure*}
\begin{figure*}
\centering
\includegraphics[width=.95\textwidth]{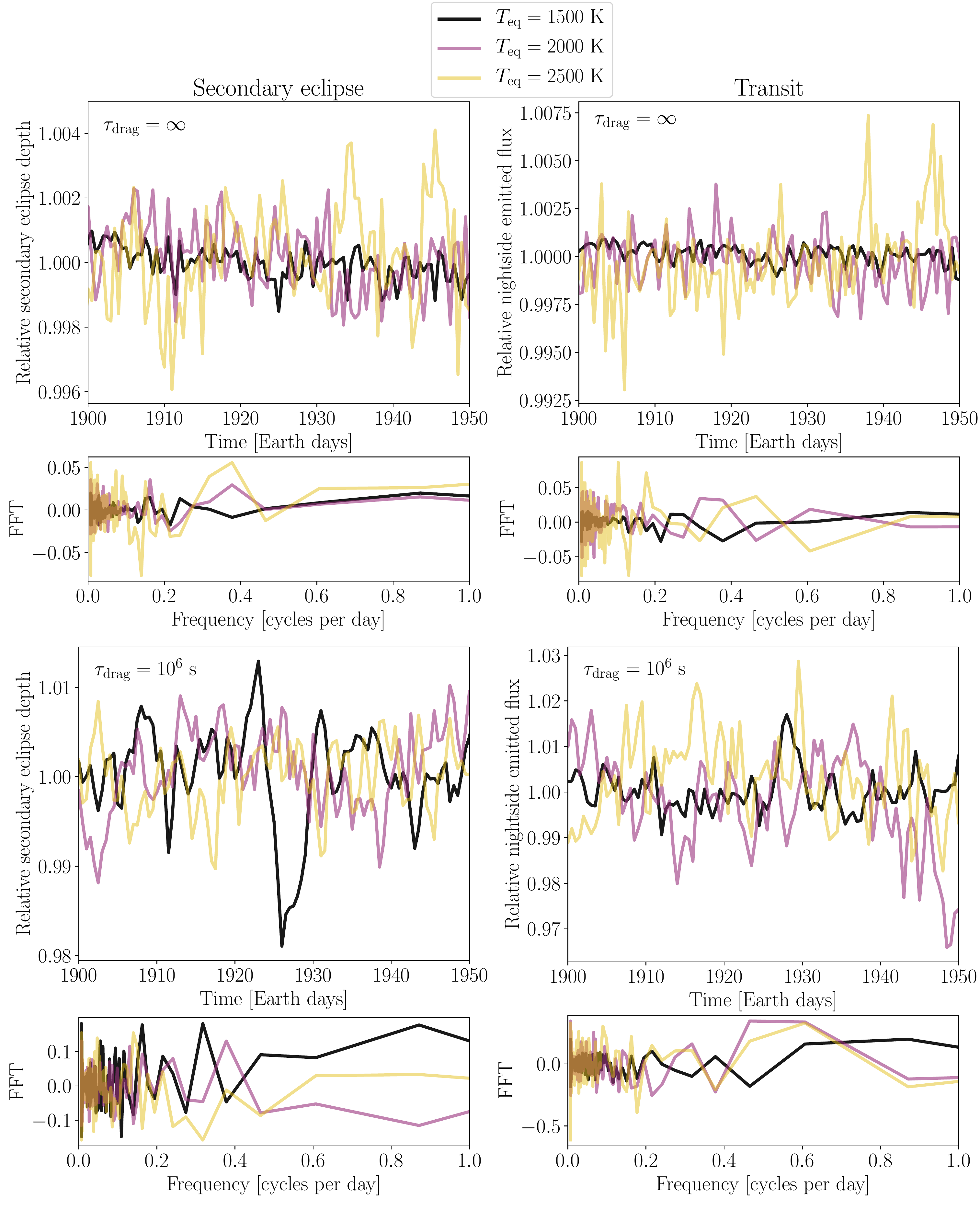}
\caption{Normalized secondary eclipse depth vs. time (left) and in-transit emitted flux vs. time for simulations with $\tau_\mathrm{drag} = \infty$ (top) and $\tau_\mathrm{drag} = 10^6~\mathrm{s}$ (bottom). Curves show results from simulations with $T_\mathrm{eq} = 1500,~2000,~\mathrm{and}~2500~\mathrm{K}$, with darker lines corresponding to cooler temperatures. Simulated data are plotted twice for every Earth day, with 50 Earth days of time shown in each plot. The fast Fourier transform of each light curve is shown below the plot. Note that here we simply output the flux variations from hemispheres centered upon the substellar point (left) and anti-stellar point (right) rather than computing more detailed mock secondary eclipse and transit spectra. In general, for the cases with $\tau_\mathrm{drag} = \infty$ the secondary eclipse depth and in-transit emitted flux amplitudes are $\lesssim 0.5\%$, though they increase with increasing equilibrium temperature. For the case with $\tau_\mathrm{drag} = 10^6~\mathrm{s}$, both amplitudes are markedly larger, at the $\sim 1\%$ level for secondary eclipse depth and few percent level for in-transit emitted flux. The Fourier transforms of the light curves show no clear peak, but in general the variability is low-frequency, with significant power at frequencies $\lesssim 0.4$ cycles per day.}
  \label{fig:sececlip_transit_var}
\end{figure*}
\begin{figure}
\centering
\includegraphics[width=.45\textwidth]{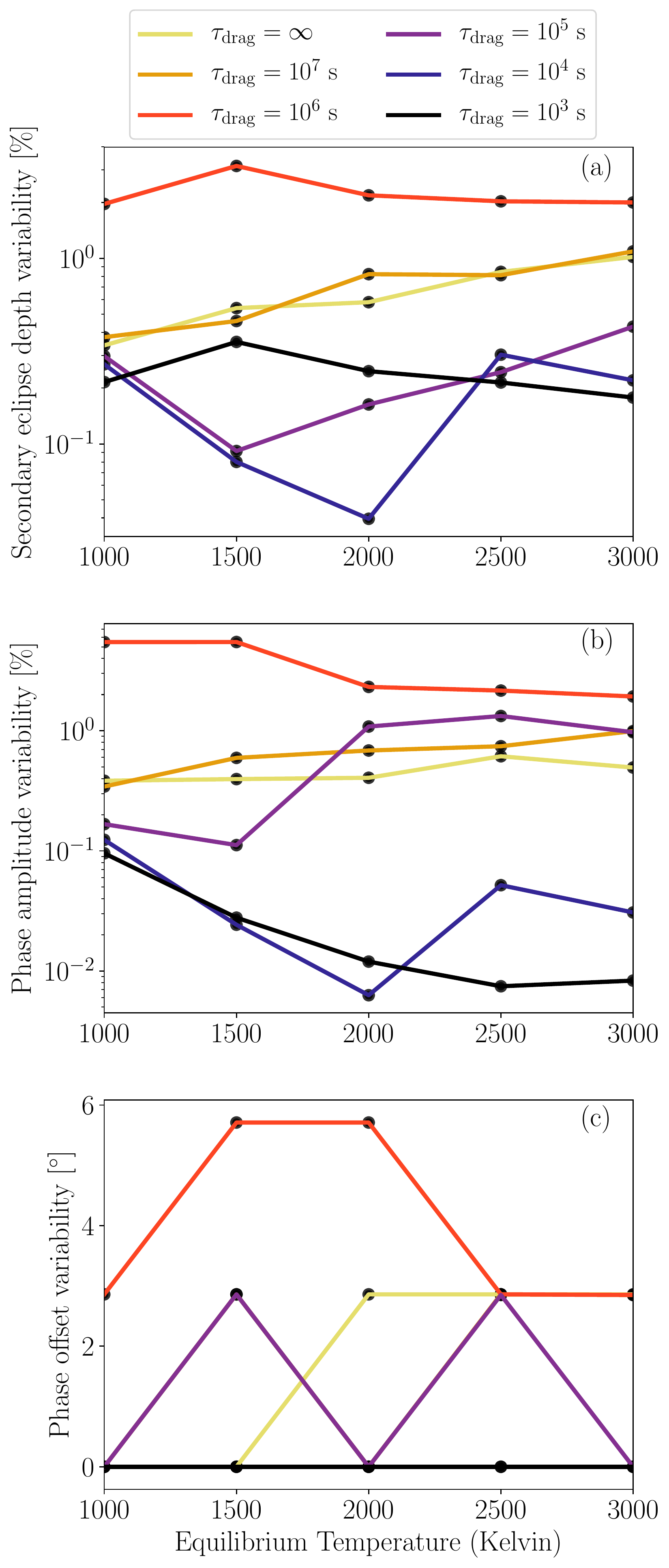}
\caption{Amplitude of variability in key light curve parameters over the last $150$ days of simulation time. (a) Percent variation in the flux from the hemisphere centered on the substellar point for simulations with varying $\tau_\mathrm{drag} = 10^3 - \infty \ \mathrm{s}$ and $T_\mathrm{eq} = 1000 - 3000 \ \mathrm{K}$. The amplitude of emergent substellar flux variation generally increases with increasing $\tau_\mathrm{drag}$, but is largest for $\tau_\mathrm{drag} = 10^6 \ \mathrm{s}$. Generally, the amplitude of variation is $\lesssim 2\%$. (b) Percent variation in the phase curve amplitude for the same set of simulations. The variation is 1-2 orders of magnitude larger for simulations with $\tau_\mathrm{drag} \geq 10^5 \ \mathrm{s}$. (c) Absolute variation in phase curve offset, in degrees longitude, for the same set of simulations. Simulations with strong drag ($\tau_\mathrm{drag} \leq 10^4 \ \mathrm{s}$) show no variation in phase curve offset, and the $\tau_\mathrm{drag} = 10^{7}~\mathrm{s}$ and $\tau_\mathrm{drag} = \infty$ case show the same variability amplitude in phase curve offset.}
  \label{fig:lightcurve_var}
\end{figure}
\\ \indent In \Fig{fig:lightcurvemaxmin}, we show the amplitude of variability in the light curve from each simulation. In general, the flux variability at any given orbital phase is small, at the $\lesssim 1-3\%$ level over a broad range in equilibrium temperature and drag strength. As we found for the time-variability of atmospheric temperature and wind speed, the variability in emitted flux increases with increasing drag timescale and increasing equilibrium temperature up to $\tau_\mathrm{drag} = 10^6~\mathrm{s}$, then decreases at even larger values of $\tau_\mathrm{drag}$. Additionally, as before the case with $\tau_\mathrm{drag} = 10^6~\mathrm{s}$ has the largest amplitude of flux variability.   \\
\indent Using these phase curves, we then calculate three key parameters: (1) the flux emerging from the hemisphere centered on the substellar point, which is the flux the planet radiates at secondary eclipse, (2) the phase curve amplitude, measured as the difference between the maximum and minimum flux of the phase curve, and (3) the phase curve offset, which is the offset in orbital phase between the phase at which the Earth-facing hemisphere of the planet is brightest and secondary eclipse. We show the resulting secondary eclipse variation versus time for a subset of simulations with $T_\mathrm{eq} = 1500,~2000,~\mathrm{and}~2500~\mathrm{K}$ and $\tau_\mathrm{drag} = 10^6~\mathrm{s}$ and $\tau_\mathrm{drag} = \infty$ in \Fig{fig:sececlip_transit_var}. Then, we calculate the normalized amplitude of variability of these three parameters for all simulations, using the same method as in \Eq{eq:rmsmax}, and plot the amplitude of time-variability in the secondary eclipse flux, phase curve amplitude, and phase offset in \Fig{fig:lightcurve_var}.
\subsubsection{Secondary eclipse depth}
\indent Figure \ref{fig:sececlip_transit_var} shows the time-evolution of our simulated secondary eclipse depths for a subset of cases with $T_\mathrm{eq} = 1500,~2000,~\mathrm{and}~2500~\mathrm{K}$ and $\tau_\mathrm{drag} = \infty$ and $\tau_\mathrm{drag} = 10^6~\mathrm{s}$. Generally, we find that the secondary eclipse depths for models with $\tau_\mathrm{drag} = \infty$ are variable at the $\lesssim 0.5\%$ level, and the secondary eclipse depths for models with $\tau_\mathrm{drag} = 10^6~\mathrm{s}$ are variable at the $\sim 1\%$ level. For the case with $\tau_\mathrm{drag} = \infty$, the secondary eclipse depth variability is markedly larger for higher $T_\mathrm{eq}$, but there is less of a difference in secondary eclipse depth with $T_\mathrm{eq}$ for the cases with $\tau_\mathrm{drag} = 10^6~\mathrm{s}$. This secondary eclipse depth variability is only quasi-periodic, with small-amplitude variability occurring on a timescale of days and larger-amplitude variability occurring on timescales of weeks to months. The Fourier transform of each light curve shows that there are not individual dominant variability frequencies. This differs from the results of \cite{Showmanetal_2009}, who found that the simulated variability in secondary eclipse depth was regular in time. 
However, the Fourier transform shows that the variability is low-frequency, with significant power at frequencies below 0.5 cycles per day. Note that the Fourier transform of the variability in phase curve amplitude and offset (not shown) shows similar power at low frequencies.  \\ 
\indent Though it may not be detectable in the near future, we also show the variability of the emitted flux from the nightside of the planet during transit in Figure \ref{fig:sececlip_transit_var}. The variability in the emitted flux from the nightside of the planet is significant, and is larger than the secondary eclipse variability for both $\tau_\mathrm{drag} = \infty$ and $\tau_\mathrm{drag} = 10^6~\mathrm{s}$ by a factor of $\sim 2$. Similar to the secondary eclipse depth variability, there is no single dominant frequency of the variability in emitted flux from the nightside hemisphere.  \\
\indent \Fig{fig:lightcurve_var}(a) shows the amplitude of the variability in secondary eclipse depth for all simulations. We find that the secondary eclipse depths of our model hot Jupiters are variable at the $\lesssim 2\%$ level. The variability in secondary eclipse depth generally increases with increasing equilibrium temperature, from $\sim 0.3\% - 1\%$ over $T_\mathrm{eq} = 1000-3000~\mathrm{K}$ for the case with no applied drag. The variability in secondary eclipse depth found in our GCM experiments is consistent with both \cite{Showmanetal_2009}, who found secondary eclipse variability of $\lesssim 1\%$, and \cite{Menou:2019aa}, who found $\lesssim 2\%$ variations in dayside flux. Similar to the temperature and wind speed variability, the case with $\tdrag = 10^6~\mathrm{s}$ shows significant variability in secondary eclipse depth.
It is notable that the cases with strong drag ($\tdrag \lesssim 10^5~\mathrm{s}$) show significantly less variability than cases with weaker drag. Additionally, there is not a strong dependence of the time-variability in secondary eclipse depth with equilibrium temperature for cases with $\tau_\mathrm{drag} \lesssim 10^4~\mathrm{s}$. We will return to our results for secondary eclipse time-variability in \Sec{sec:disctimevar} to discuss how secondary eclipse variability may be observable with \textit{JWST}.
\subsubsection{Phase curve amplitude}
\Fig{fig:lightcurve_var}(b) shows the variability in phase curve amplitude for our grid of simulations. The variability in phase curve amplitude is comparable to the variability in secondary eclipse depth, as the variability in phase curve amplitude is dominated by the variability in the hottest hemisphere. In general, for simulations with $\tau_\mathrm{drag} \gtrsim 10^5~\mathrm{s}$ we find that the phase curve amplitude variability is $\sim 0.3\% - 3\%$. However, for strong $\tau_\mathrm{drag} \lesssim 10^4~\mathrm{s}$ the phase curve amplitude is much less variable, at the $\lesssim 0.1\%$ level for $T_\mathrm{eq} = 1000 \ \mathrm{K}$ and as small as $0.01\%$ for $T_\mathrm{eq} \gtrsim 2000~\mathrm{K}$. Additionally, unlike the secondary eclipse depth, the phase curve amplitude variability is still significant for the case with $\tdrag = 10^5~\mathrm{s}$. This may be related to how the case with $\tdrag = 10^5~\mathrm{s}$ lies in an intermediate regime between an atmosphere dominated by a superrotating jet and an atmosphere dominated by dayside-to-nightside flow (see Figure 5 of \citealp{Komacek:2017}). As a result, the superrotating equatorial jet is slowed, but not as much as it is with stronger drag. This may lead to an oscillation due to the similarity between the wave driving and damping timescales (which are both $\sim 10^5~\mathrm{s}$). This residual variability in the wave pattern is seen through the phase curve amplitude variability. 
\subsubsection{Phase curve offset}
\Fig{fig:lightcurve_var}(c) shows the variability in phase curve offset for our suite of simulations. For cases with weak drag ($\tau_\mathrm{drag} \gtrsim 10^5~\mathrm{s}$), the variability in phase offset is significant, up to $\approx 5^{\circ}$. For the cases with very weak drag ($\tdrag \gtrsim 10^7~\mathrm{s}$), the phase curve offset variability reaches $\approx 3^{\circ}$ at $T_\mathrm{eq} = 3000~\mathrm{K}$. As we found for all other observable properties, the $\tdrag = 10^6~\mathrm{s}$ case shows the largest variability in phase curve offset, up to $\approx 5^{\circ}$. We do not find significant variability for the cases with strong drag ($\tau_\mathrm{drag} \lesssim 10^4~\mathrm{s}$). This is both because these cases with strong drag do not have a significant phase offset to begin with due to their weak wind speeds and because the strong drag then damps any potential variability in the phase curve offset.
\section{Discussion}
\label{sec:disctimevar}
\subsection{Mechanism of the variability}
\label{sec:mechanism}
\begin{figure*}
\begin{center}
\includegraphics[width=1\textwidth]{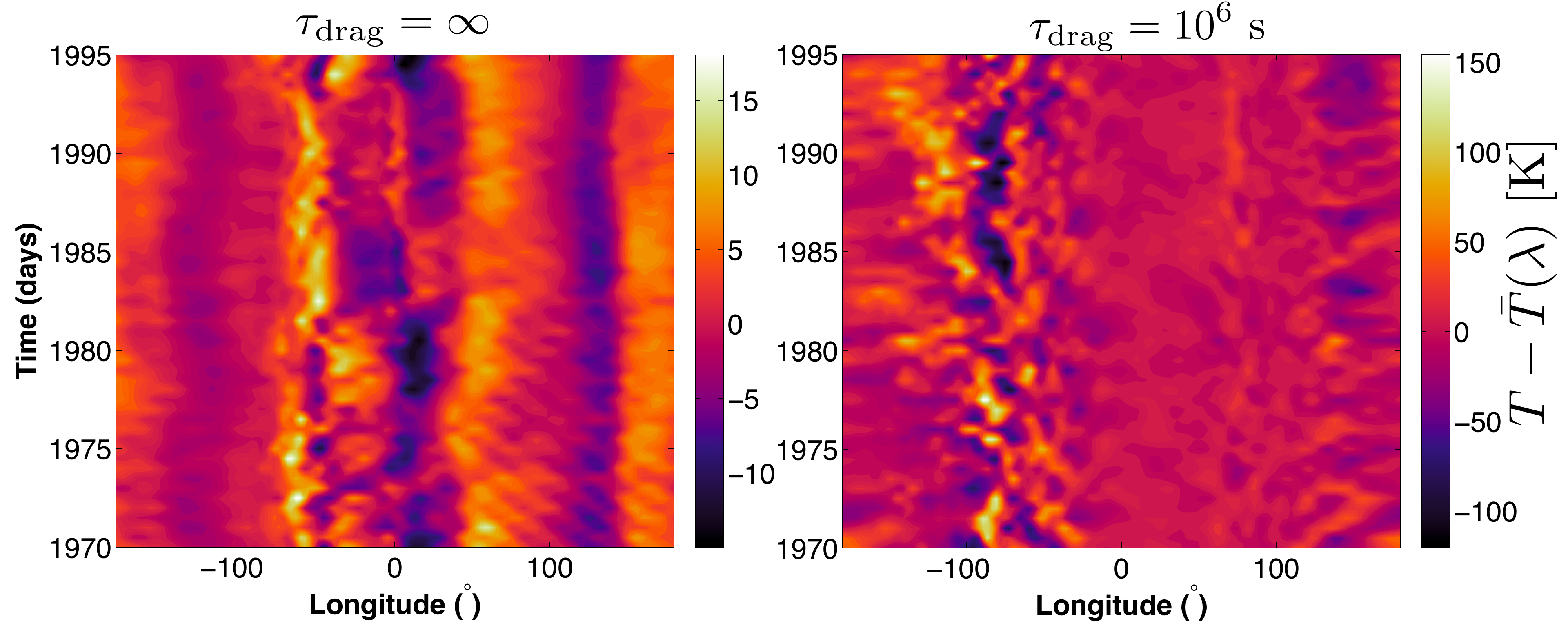}
\caption{Hovm\"{o}ller diagrams of temperature deviations from the time-averaged temperature, averaged over latitudes of $\pm 15^{\circ}$, as a function of longitude ($\lambda$) at a given time at a pressure of $80~\mathrm{mbar}$ for simulations with $T_\mathrm{eq} = 1500~\mathrm{K}$ and $\tau_\mathrm{drag} = 10^6~\mathrm{s}~\mathrm{and}~\infty$. The diagrams show only 25 days of simulation time, from $t = 1970 - 1995~\mathrm{days}$. Multiple wave trains generated on the western hemisphere of the dayside and propagating eastward are visible in the $\tau_\mathrm{drag} = \infty$ panel. Additionally, westward propagation of waves is visible near the eastern terminator. The characteristic variability for the case with $\tau_\mathrm{drag} = 10^6~\mathrm{s}$ is an order of magnitude larger than for the case with $\tau_\mathrm{drag} = \infty$, and this variability is much noisier. As a result, it is difficult to visually discern wave propagation for this case, as expected from the time-series analysis of secondary eclipse depth shown in \Fig{fig:sececlip_transit_var}.}
  \label{fig:Tlongtimefocus}
 \end{center}
\end{figure*}
\indent Though in this work we do not perform a detailed analysis of the simulated time-variability, in this section we summarize some of the key features of the variability and speculate on possible mechanisms that cause this variability. Recall from the temperature variability maps in \Fig{fig:temp_maps} that the bulk of the time-variability is confined to the equatorial regions, within a latitudinal band between $\pm 15^{\circ}$. Additionally, the variability is largest near both the western and eastern terminators. Notably, the variability at the western terminator coincides with the position of a large temperature contrast between the cool nightside and hot dayside. \\
\indent To analyze the time-variability in greater detail, we return to our analysis of how the equatorial temperature pattern changes as a function of time. \Fig{fig:Tlongtimefocus} shows Hovm\"{o}ller diagrams for the equatorial temperature deviation from the time-averaged temperature as a function of longitude for simulations with $T_\mathrm{eq} = 1500~\mathrm{K}$ and $\tau_\mathrm{drag} = 10^6~\mathrm{s}~\mathrm{and}~\infty$. This is similar to \Fig{fig:Tlongtime}, but here the temperature deviation is relative to the local time-average of temperature instead of the longitude-average. Additionally, we have focused on only 25 days of the simulation, in order to visually discern features of the time-variability. One can see that the amplitude of variability for the case with $\tau_\mathrm{drag} = 10^6~\mathrm{s}$ is generally an order of magnitude larger than the case with $\tau_\mathrm{drag} = \infty$. Additionally, the spatial distribution of such variability is somewhat different. The strongest variability for the case with $\tau_\mathrm{drag} = 10^6~\mathrm{s}$ occurs near the western terminator, with additional variability eastward of the eastern terminator where there is vertical advection due to the Gill pattern. However, the strongest variability for the case with $\tau_\mathrm{drag} = \infty$ occurs just eastward of the western terminator. This variability occurs over a much smaller range of longitudes than in the case with $\tau_\mathrm{drag} = 10^6~\mathrm{s}$, likely due to the much lower-amplitude variability in the case with $\tau_\mathrm{drag} = \infty$.  In general, the variability for the case with $\tau_\mathrm{drag} = 10^6~\mathrm{s}$ is fairly noisy, with individual wave trains difficult to see. However, the case with $\tau_\mathrm{drag} = \infty$ does have clear wave propagation, as we discuss next.  \\
\indent Now we speculate on the wave generation mechanism using the results shown in \Fig{fig:Tlongtimefocus} for the case with $\tau_\mathrm{drag} = \infty$. The temperature pattern appears to have tilts in longitude-time phase space such that the tilts are eastward (with increasing time) eastward of the western terminator and westward (again, with increasing time) westward of the western terminator. Overall, the variability is relatively noisy, with periodicity difficult to discern. However, waves can be seen over a broad range of longitudes from $-50^\circ$ to $180^\circ$. Both eastward and westward propagating modes can be prominently seen, superimposed together. The eastward-propagating waves propagate from their generation region on the western half of the dayside to the substellar point in a timespan of a few days, corresponding to relatively low phase speeds of $\sim 600~\mathrm{m}~\mathrm{s}^{-1}$. 
The westward propagating waves that occur at eastward longitudes have similar phase speeds. The basic characteristics of the eastward-propagating waves are reminiscent of Kelvin waves \citep{Holton:2013}: they propagate eastward, are confined to equatorial regions, and have relatively low phase speeds. Additionally, one might expect that the westward propagating branch of waves may be equatorial Rossby waves, which are also planetary-scale waves with relatively low phase speeds that propagate westward. \\
\indent To examine the wave dynamics in more detail, we compare the $\sim 600~\mathrm{m}~\mathrm{s}^{-1}$ propagation speed of both the eastward and westward propagating waves to that of Rossby and Kelvin waves. This propagation speed is broadly consistent with the expected phase speed of Rossby waves\footnote{The phase speed of a Rossby wave in the limit of zero background wind is approximately $c \approx -\beta/k^2$, where $\beta$ is the latitudinal gradient in the Coriolis parameter and $k$ is horizontal wavenumber. Using a planetary radius and rotation rate appropriate for HD 209458b, a planetary wavenumber of 16 (estimated from the case with $\tau_\mathrm{drag} = \infty$), and evaluating the phase speed at the equator, we find $c \approx - 600~\mathrm{m}~\mathrm{s}^{-1}$.}, but is a factor of two smaller than that of Kelvin waves\footnote{In the limit of long vertical wavelength, the phase speed of a Kelvin wave without background flow is $c \approx 2NH$, where $N$ is the Brunt-Vaisala frequency and $H$ is the scale height \citep{showman_2013_doppler}. In the vertically isothermal limit, $c \approx R\sqrt{T/c_p}$, where $R$ is the specific gas constant, $T$ is temperature, and $c_p$ is the specific heat. For parameters appropriate for the simulation with $T_\mathrm{eq} = 1500~\mathrm{K}$, $c \approx 1250~\mathrm{m}~\mathrm{s}^{-1}$.} when ignoring the Doppler shift due to the background jet. Including the Doppler shift of the background jet (with a zonal-mean equatorial speed of $\sim 4000~\mathrm{m}~\mathrm{s}^{-1}$), one would expect the phase speeds of both the Doppler-shifted Kelvin and Rossby waves to be an order of magnitude larger than that found in \Fig{fig:Tlongtimefocus}. As a result, it is unclear whether Rossby and/or Kelvin waves are the cause of the equatorial variability in our simulations. An alternative is that the equatorial waves are composed of mixed Rossby-gravity modes, which is consistent with the anti-symmetry of the temperature perturbations across the equator found in \Fig{fig:temp_maps}. Most likely, the equatorial time-variability in our simulations is caused by a combination of mixed Rossby-gravity, Kelvin, and Rossby waves, which cannot be discerned without more detailed wavenumber-frequency analysis. \\
\indent Note that the 
$\sim 20~\mathrm{day}$ longitudinal propagation timescale (across the circumference of the planet) of the waves is not dissimilar to the $\sim 40~\mathrm{day}$ periodicity of variability seen by \cite{Showmanetal_2009}, given the stark differences in our modeling approaches. With higher-resolution simulations, it may be possible to do a more detailed analysis of the wavenumber-frequency spectrum of hot Jupiter flows, akin to that in the Earth tropical dynamic literature \citep{Wheeler:1999}. Such an analysis has been performed on simulations of brown dwarfs and Jupiter-like planets \citep{Showman:2019aa,Young:2019aa}. A similar analysis is outside the scope of this work, but would be beneficial for determining the wave types causing time-variability in hot Jupiter atmospheres.  \\
\indent Our analysis of time-variability shows that the maximum amplitude of variability in both intrinsic and observable properties generally occurs at an intermediate drag timescale of $\tau_\mathrm{drag} = 10^6~\mathrm{s}$. This drag timescale corresponds to a transition in the flow pattern from a zonal jet at weaker drag ($\tau_\mathrm{drag} > 10^6~\mathrm{s}$) and a standing Matsuno-Gill pattern or day-to-night flow at stronger drag ($\tau_\mathrm{drag} < 10^6~\mathrm{s}$) \citep{Komacek:2015,Komacek:2019aa}. We propose that the basic-state flow structure is more dynamically unstable in the $\tau_\mathrm{drag} = 10^6~\mathrm{s}$ case than with either stronger or weaker drag. This enhanced instability is evident from the hemispheric asymmetry of the flow pattern for cases with $\tau_\mathrm{drag} = 10^6~\mathrm{s}$ and hot equilibrium temperatures in excess of 1500 K (see figure 5 of \citealp{Komacek:2017}). The enhanced instability of the flow at intermediate $\tau_\mathrm{drag}$ is a natural consequence of the changing properties of the Matsuno-Gill pattern and equatorial jet. The Matsuno-Gill pattern is weak with very strong drag, and gets stronger with increasing $\tau_\mathrm{drag}$ \citep{Showman_Polvani_2011}. As the Matsuno-Gill pattern can become unstable \citep{Showman:2010}, the increasing strength of the Matsuno-Gill pattern will cause the flow to be more unstable and exhibit greater variability with increasing $\tau_\mathrm{drag}$. However, for weak drag with $\tau_\mathrm{drag} > 10^6~\mathrm{s}$ the zonal jet becomes strong \citep{Komacek:2017}, which likely weakens the Matsuno-Gill pattern. Thus, for very weak drag the instability of the Matsuno-Gill pattern should also weaken, leading to increased instability and enhanced time-variability at an intermediate value of $\tau_\mathrm{drag} = 10^6~\mathrm{s}$. The development of the strong eastward jet at $\tau_\mathrm{drag} > 10^6~\mathrm{s}$ also implies an increase in the meridional gradient of potential vorticity at low latitudes, which could potentially act to help stabilize the Matsuno-Gill pattern by making it less likely for the eddy structure associated with the Matsuno-Gill pattern to exhibit reversals in the potential vorticity gradient.  Future work will be needed to test these and other hypotheses for the local maximum in variability near $\tau_\mathrm{drag} = 10^6~\mathrm{s}$. \\
\indent Lastly, we speculate on the process generating these planetary-scale waves. It appears that the waves originate near the western terminator, at the location where the temperature increases sharply going from nightside to dayside. This potential wave generation region corresponds to a hydraulic jump or acoustic shock on the western limb that has been noted previously by many authors (e.g., \citealp{Showmanetal_2009,Heng:2012a,perna_2012,Fromang:2016}). Note that though our numerical simulations do not explicitly resolve acoustic shocks, the hydraulic jumps seen in this work and that of \cite{Showmanetal_2009} are analogous features. If they are present, shocks may then act to cause stirring that generates planetary-scale propagating waves. For example, small fluctuations in the longitude of the shock over time would act as an impulse that excites Kelvin waves which would then propagate eastward away from the shock feature. Future work building upon that of \cite{Fromang:2016} using simulations that allow for shocks and the resulting wave generation can shed further light on the cause of time-variability in hot Jupiter atmospheres. 
\subsection{Doppler wind speeds: effects of varying equilibrium temperature and drag timescale}
\label{sec:dopplertrends}
\begin{figure*}
\begin{center}
\includegraphics[width=1\textwidth]{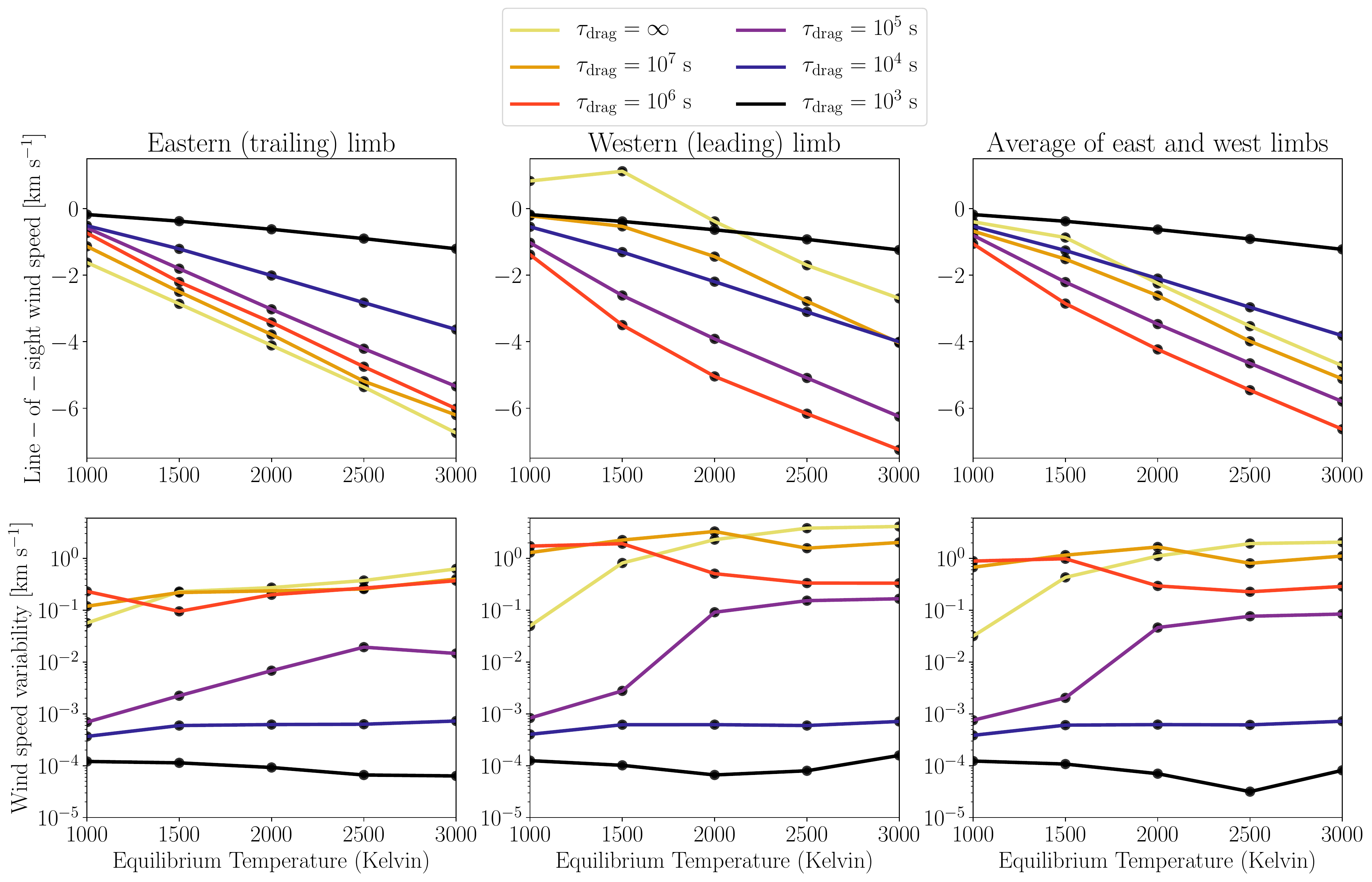}
\caption[Simulated line-of-sight wind speed at a pressure of $1~\mathrm{mbar}$ (top) and the maximum time-variability in these wind speeds over the last 150 days of the simulation (bottom).]{Simulated line-of-sight wind speed at a pressure of $1~\mathrm{mbar}$ (top) and the maximum time-variability in these wind speeds over the last 150 days of the simulation (bottom). Specifically, the top three panels show the time-mean over the last 150 days of the average wind speed toward the observer across the eastern and western terminator and their average. The bottom three panels show the maximum variability in these average wind speeds (i.e., the difference between the maximum and minimum line-of-sight wind speeds) over the last 150 days of simulation time. The top panels share a linear scale, with positive wind speeds corresponding to redshifts (motion away from the observer), while negative wind speeds correspond to blueshifts (motion toward the observer). The bottom panels share a logarithmic scale. Wind speeds at the eastern limb are always negative, and increase in absolute magnitude with increasing equilibrium temperature and decreasing drag strength. For simulations with high equilibrium temperature and strong drag, wind speeds at the western limb are normally negative, but do not monotonically increase in absolute magnitude with decreasing drag strength. The average of the wind speed on the east and west limbs is always negative and increases in absolute magnitude with increasing equilibrium temperature. The variability in terminator wind speeds is large for cases with weak drag ($\tau_\mathrm{drag} \ge 10^6~\mathrm{s}$) and is normally larger on the western limb than on the eastern limb. The variability in terminator-averaged wind speeds can be a few kilometers per second for cases with weak drag ($\tau_\mathrm{drag} \ge 10^6~\mathrm{s}$), potentially observable with current high-resolution spectroscopic techniques.}
  \label{fig:udoppler}
 \end{center}
\end{figure*}
High-resolution spectroscopy has constrained the terminator-averaged blueshift for HD 209458b \citep{Snellen:2010} and both the terminator-averaged and the terminator-separated (i.e., separated into those for the eastern and western limbs) Doppler shifts for HD 189733b \citep{Louden:2015,Wyttenbach:2015,Brogi:2015}. These observations probe relatively high altitudes in the atmosphere, corresponding to pressures of $\sim 1~\mathrm{mbar}$ or less. Under the assumption that HD 189733b is tidally locked to its host star, \cite{Louden:2015} separated the Doppler shifts on the leading and trailing limbs of the planet during a transit event. \citeauthor{Louden:2015} found that the wind speeds were redshifted at the leading (western) limb and blueshifted at the trailing (eastern) limb, exactly what is expected from the presence of a superrotating jet \citep{showman_2002,Showman_Polvani_2011,Kempton:2012aa,showman_2013_doppler,Zhang:2017b,Flowers:2018aa}.  \\
\indent In \Fig{fig:udoppler} (top), we show the simulated line-of-sight wind speeds at a pressure of $1~\mathrm{mbar}$ at both the eastern and western terminators and their average for each of our simulations varying equilibrium temperature and drag strength. Specifically, we show the time-mean of the latitudinally-averaged zonal wind speeds on the terminators. We define positive wind speeds as away from the observer (i.e., westward on the eastern limb and eastward on the western limb) and negative wind speeds as toward the observer (i.e., eastward on the eastern limb and westward on the western limb). \\
\indent We find that the wind speeds on the eastern limb are always negative (toward the observer) and increase in absolute magnitude with increasing equilibrium temperature and decreasing drag timescale, as expected from the simulated net Doppler wind speeds of previous studies \citep{Kempton:2012aa,showman_2013_doppler,Kempton:2014}. The wind speeds on the western limb are negative for all simulations with $\tau_\mathrm{drag} \le 10^7~\mathrm{s}$ and generally increase in magnitude with increasing equilibrium temperature. On the western limb, models with no drag experience positive winds (i.e., away from the observer leading to a Doppler redshift) for modest $T_\mathrm{eq}$ of 1000-1500 K.  This is the regime where the equatorial jet is most coherent with longitude (see \Fig{fig:temp_maps}d, and Figure 5 of \citealp{Komacek:2017}) and thus this is the signature of the equatorial jet blowing away from observer. Note that the wind speeds on the western limb do not change monotonically with drag strength, as the fastest winds at the western limb are for the simulations with intermediate drag strengths. The average of the two limbs is always negative, with the fastest winds toward the observer for simulations with $\tau_\mathrm{drag} = 10^6~\mathrm{s}$. \\
\indent We note that our simulations with $\tau_\mathrm{drag} = \infty$ are consistent with the observations of \cite{Louden:2015} in that at the low equilibrium temperatures of 1000-1500 K relevant for HD 189733b, the eastern limb shows a net blueshift, the western limb a net redshift, and their combination shows a net blueshift (see their Figure 3). Our models predict that hotter hot Jupiters may have a western limb that (like the eastern limb) becomes blueshifted.   
\subsection{Future prospects for observing time-variability}
\subsubsection{Light curves}
\indent As shown in \Sec{sec:resultstimevarobs}, we expect that the secondary eclipse depths of hot Jupiters are time-variable at the $\lesssim 2\%$ level. Notably, we show in \Fig{fig:lightcurve_var} that for atmospheres with weak drag ($\tau_\mathrm{drag} \gtrsim 10^7~\mathrm{s}$), the secondary eclipse variability increases with increasing equilibrium temperature, from $\sim 0.3\%$ at $T_\mathrm{eq} = 1000~\mathrm{K}$ to $\sim 1\%$ at $T_\mathrm{eq} = 3000~\mathrm{K}$. This variability would not be detectable with \textit{Spitzer} given current upper limits \citep{Agol:2010,Kilpatrick:2019aa}. As a result, we next determine if the observable variability from our simulated hot Jupiters would be detectable with \textit{JWST}. \\
\indent For \textit{JWST}, we assume a pessimistic noise floor of $\approx 30~\mathrm{ppm}$ with NIRCam \citep{Greene:2015} and consider a planet-to-star flux ratio of $\approx 0.002$ in this wavelength range (relevant for an HD 209458b-like planet with $T_\mathrm{eq} = 1500~\mathrm{K}$). With the above assumptions, we estimate that \textit{JWST} will be able to detect planetary flux variability on the order of $\gtrsim 1.5\%$. Assuming that the planet to star flux ratio increases as a blackbody, we expect that for hot Jupiters with $T_\mathrm{eq} \gtrsim 2000~\mathrm{K}$ time-variability in secondary eclipse depth in the near-infrared on the order of $\gtrsim 0.1\%$ will likely be detectable. Our simulations show that variability amplitudes that could be observable with \textit{JWST} are possible for planets with weak $\tau_\mathrm{drag} \gtrsim 10^6~\mathrm{s}$.  \\
\indent We expect potentially significant variability in the phase curve offset, up to $3^{\circ}$ for the hottest hot Jupiters without atmospheric drag and potentially up to $\approx 5^{\circ}$ for hot Jupiters with moderate $\tdrag =10^5 - 10^6~\mathrm{s}$. Though the variability we model is generally not as large as the variability that \cite{Mooij2016} and \cite{Jackson:2019aa} found for HAT-P-7b and Kepler-76b in visible wavelengths and \cite{Bell:2019aa} found in the infrared for WASP-12b, the variability in phase curve offset that we find is still signifiant and potentially observable with multiple phase curves. Additionally, note that our simulations do not include magnetic effects, which may greatly affect the phase curve offset at temperatures $\gtrsim 1500~\mathrm{K}$, and we do not include clouds (see \Sec{sec:other} for further discussion). Note that though full-phase light curves are a very time-intensive measurement \citep{Crossfield:2015,Parmentier:2017}, it is possible that future dedicated missions (e.g., \textit{ARIEL}) may be able to observe multiple phase curves of the same object in order to constrain atmospheric time-variability.
\subsubsection{Doppler wind speeds}
\begin{figure}
\begin{center}
\includegraphics[width=0.5\textwidth]{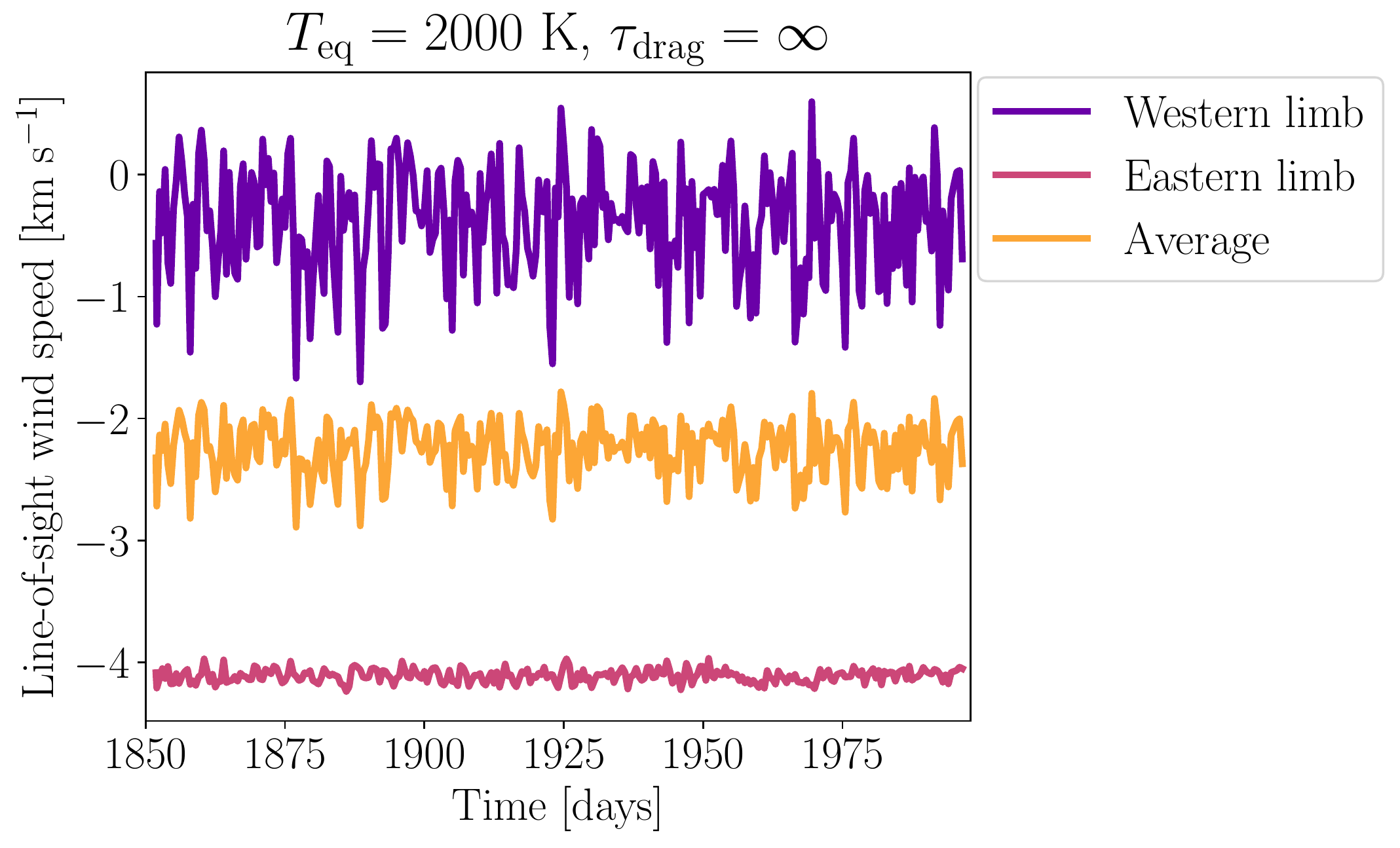}
\caption{Time-dependence of the line-of-sight wind speeds at $1~\mathrm{mbar}$ from the simulation with $T_\mathrm{eq} = 2000~\mathrm{K}$ and $\tau_\mathrm{drag} = \infty$. The Doppler shift at the western limb can reverse sign from a blueshift to a redshift over short timescales. The amplitude of variability is much smaller on the eastern limb, which is always blueshifted. As a result, the terminator-averaged wind speeds are always blueshifted. Three other GCM experiments that we performed show similar behavior, with the Doppler shift at the western limb reversing sign. High-resolution spectroscopic techniques which measure the Doppler shift at the ingress and egress separately may find a change in sign of the wind speed at the leading (western) limb due to purely hydrodynamic effects.}
  \label{fig:udoppler_time}
 \end{center}
\end{figure}
\indent It is possible that the Doppler wind speeds presented in \Sec{sec:dopplertrends} may vary in time, providing information about the variability of the underlying atmospheric circulation. In \Fig{fig:udoppler} (bottom), we show the amplitude of time-variability in the line-of-sight wind speeds at the eastern and western terminators, along with their sum. We find that the variability is orders-of-magnitude larger for simulations with $\tau_\mathrm{drag} \ge 10^6~\mathrm{s}$ than for those with $\tau_\mathrm{drag} \le 10^4~\mathrm{s}$. Generally, for simulations with $\tau_\mathrm{drag} \ge 10^6~\mathrm{s}$ the wind speed variability on the eastern limb is smaller than that on the western limb. \\
\indent \Fig{fig:udoppler_time} shows that for the case with $T_\mathrm{eq} = 2000 ~\mathrm{K}$ and no drag, the line-of-sight wind speed averaged over the western limb at a pressure of $1~\mathrm{mbar}$ can reverse from blueshifted to redshifted over short timescales. This reversal of the line-of-sight wind speed at the western limb occurs in 4 separate experiments, all with weak drag. The specific experiments that show this behavior have equilibrium temperatures and drag timescales of $T_\mathrm{eq} = 1000~\mathrm{K}$ and $\tau_\mathrm{drag} = 10^7~\mathrm{s}$,  $T_\mathrm{eq} = 1500~\mathrm{K}$ and $\tau_\mathrm{drag} = 10^7~\mathrm{s}$, $T_\mathrm{eq} = 2000 ~\mathrm{K}$ and $\tau_\mathrm{drag} = \infty$, and $T_\mathrm{eq} = 2500~\mathrm{K}$ and  $\tau_\mathrm{drag} = \infty$. This behavior occurs because the wind speed at the western limb is on average near zero, due to a cancellation between the eastward (blueshifted) superrotating jet and westward (redshifted) day-to-night flow at higher latitudes. As a result, similar-amplitude variability in the wind speed as found in other simulations can lead to a reversal in the line-of-sight wind direction.   \\
\indent For all simulations with $T_\mathrm{eq} \ge 1500~\mathrm{K}$ and $\tau_\mathrm{drag} \ge 10^7~\mathrm{s}$, the wind-speed variability on the western limb and for the average of the eastern and western limb can be $\gtrsim 1~\mathrm{km}~\mathrm{s}$. This variability is both significant relative to the wind speeds themselves and may be observable, given that the current state-of-the-art uncertainty on limb-separated wind speeds is $\approx \pm 1.5~\mathrm{km}~\mathrm{s}^{-1}$, with the uncertainty for the limb-averaged wind speeds $\approx \pm 0.5~\mathrm{km}~\mathrm{s}^{-1}$ \citep{Louden:2015}. 
As a result, we expect that future observations of multiple transit events with high-resolution spectroscopic techniques may enable the detection of time-variability of Doppler shifts in high-resolution spectra. 
\subsection{Other mechanisms to induce atmospheric variability}
\label{sec:other}
In this work, we undertook a first exploration of how time-variability in hot Jupiter atmospheres varies with planetary parameters using a simplified three-dimensional GCM. Notably, we did not include non-grey radiative transfer nor the effects of clouds, the latter of which is known to greatly affect phase curves and emergent spectra of hot Jupiters \citep{Demory_2013,Mooij2016,Heng:2016,Lee:2016,Parmentier:2015,Stevenson:2016,Lines:2018}. Using a GCM with passive tracers, \cite{parmentier_2013} showed that the tracer abundance can vary significantly more than the temperature field itself. As a result, this could allow for significant variability in the cloud distribution in a GCM that includes clouds. Additionally, it is possible that small variations in temperature could lead to large variations in the saturation vapor pressure of condensible species via the Clausius-Clapeyron relation (\citealp{Pierrehumbert:2010}, Ch. 2.6). This implies that small temperature variations could cause large variations in condensible vapor abundances, which could in turn cause significant cloud variability. Such cloud variability would then alter the three-dimensional structure of radiative heating and cooling, which would then feed back into the temperature structure and perhaps allow for enhanced variability in the temperature field itself relative to what our idealized models in this work show. As a result, there is significant research yet to be done into how clouds affect time-variability in hot Jupiter atmospheres, including both their intrinsic time variability and the observational consequences of this variability in spectra and phase curves. \\
\indent As discussed in \Sec{sec:introvar}, magnetic effects can have a significant effect on emergent properties of hot Jupiters. Most importantly, magnetic effects can cause the equatorial jet to reverse from eastward to westward, oscillating in time \citep{Rogers:2014,Rogers:2017}.  This would cause resulting oscillations in the sign of the phase curve offset and could cause the Doppler shifts in transit ingress and egress to reverse sign \citep{Louden:2015}. Note that we found that in certain parameter regimes purely hydrodynamic variability can also cause a reversal of the sign of the Doppler shift at the western limb (see \Fig{fig:udoppler_time}), so further work is needed to predict the impact of magnetic effects on high-resolution transmission spectra. In this work, we did not consider time-variability due to magnetic effects, though it is expected that for hot Jupiters with $T_\mathrm{eq} \gtrsim 1400~\mathrm{K}$ magnetic effects will be important \citep{Menou:2012fu,Rogers:2020,Rogers:2014}.  \\
\indent Additionally, our simulations do not capture non-hydrostatic instabilities associated with vertical shear of the horizontal winds, which have been shown to cause time-variability in the speed of the equatorial jet \citep{Fromang:2016}. As a result, our predictions for the variability in wind speeds are lower limits, and it is possible that globally averaged wind speed variability at levels of $> 10\%$ is achievable. In general, we did not include any parameterization for the drag timescale at near-photospheric levels changing as a function of time and space, as it could in an atmosphere with magnetic effects and/or vertical shear instabilities. We hope that this paper motivates future work exploring how more realistic drag parameterizations induce atmospheric time-variability in hot Jupiters. 
\section{Conclusions}
\label{sec:conclusionstimevar}
We have shown that hot Jupiter atmospheres are time-variable at potentially observable levels, even when ignoring the potential variability due to magnetic effects, vertical shear instabilities, and clouds. 
Below, we summarize the key ways in which this time-variability may be manifest in observations, along with our predictions for how terminator wind speeds vary with planetary parameters.
\begin{enumerate}
\item Hot Jupiter atmospheres are time-variable at the $\sim 0.1-1\%$ level in global-average, dayside-average, and nightside-average temperature and at the $\sim 1-10\%$ level in global-average wind speeds. This variability depends strongly on pressure, with larger-amplitude variability generally occurring at lower pressures. With repeated transit observations using high-resolution spectrographs, the $\gtrsim 1~\mathrm{km}~\mathrm{s}^{-1}$ variability in Doppler wind speeds at the terminator predicted from our simulations with $\tau_\mathrm{drag} \ge 10^6~\mathrm{s}$ may be observationally detectable.
\item We calculate model Doppler wind speeds at a pressure of $1~\mathrm{mbar}$ for the western (leading) and eastern (trailing) limbs in our suite of GCMs. We find that the eastern limb always shows a Doppler blueshift due to winds. Meanwhile, the western limb shows a small redshift in simulations without drag and with equilibrium temperatures $\le 1500~\mathrm{K}$, but shows a blueshift in most simulations with stronger drag and/or larger values of incident stellar flux. Additionally, the time-variability in Doppler wind speeds can lead to redshift-to-blueshift oscillations on the western limb. As a result, even hot Jupiters with a superrotating equatorial jet may not show a Doppler wind signature with an eastern limb blueshift and western limb redshift if they have large equilibrium temperatures and/or have significant atmospheric drag. 
\item Without considering magnetic effects and vertical shear instabilities, we expect that the secondary eclipse depths of hot Jupiters are variable at the $\lesssim 2\%$ level. For atmospheres without strong applied drag (with $\tau_\mathrm{drag} \gtrsim 10^7 \ \mathrm{s}$), the secondary eclipse depth variability increases with increasing equilibrium temperature, from $\sim 0.3\% - 1\%$ as $T_\mathrm{eq}$ increases from $1000 - 3000~\mathrm{K}$. For atmospheres with strong drag (with $\tau_\mathrm{drag} \lesssim 10^5 \ \mathrm{s}$), the secondary eclipse depth variability is always $\lesssim 0.3\%$. Variability in the secondary eclipse depth of hot planets with weak atmospheric drag is potentially observable with \textit{JWST}.
\item We expect that hot Jupiter phase curves are also significantly time-variable. Most notably, we expect that the phase offset, which has already been shown to be variable from both optical and infrared phase curves, should be variable in the infrared if the characteristic drag timescale $\tau_\mathrm{drag} \gtrsim 10^5 \ \mathrm{s}$. This variability in phase offset increases with increasing equilibrium temperature, and can be $\approx 3-5^{\circ}$ for the hottest planets with weak atmospheric drag. We also predict that the phase curve amplitude is variable at the $\sim 0.1\%-1\%$ level for atmospheres with $\tau_\mathrm{drag} \gtrsim 10^5 \ \mathrm{s}$, with this variability in phase curve amplitude largely independent of equilibrium temperature. 
\end{enumerate}

\acknowledgements
We thank the referee for helpful comments that improved the manuscript. T.D.K. acknowledges funding from the 51 Pegasi b Fellowship in Planetary Astronomy sponsored by the Heising-Simons Foundation. 
\if\bibinc n
\bibliography{References_all}
\fi

\if\bibinc y

\fi

\end{document}